\documentclass[onecolumn, numberedappendix]{aastex6}
\usepackage{color}

\newcommand{\eg}{{\it e.g.}}
\newcommand{\etal}{et~al.}

\newcommand{\ie}{{\it i.e.}}
\newcommand{\ks}{$K_{\rm s}$}
\newcommand{\vmk}{$(V-K_{\rm s})_0$}

\newcommand{\msun}{M$_{\sun}$}

\newcommand{\teff}{$T_{\rm eff}$}
\newcommand{\vsini}{$v \sin i$}
\newcommand{\mum}{$\mu$m}

\begin{document}

\title{Rotation in the Pleiades with K2: II. Multi-Period Stars }

\slugcomment{Version from \today}

\author{L.~M.~Rebull\altaffilmark{1,2},
J.~R.~Stauffer\altaffilmark{2},
J.~Bouvier\altaffilmark{3}, 
A.~M.~Cody\altaffilmark{4},
L.~A.~Hillenbrand\altaffilmark{5}, 
D.~R.~Soderblom\altaffilmark{6}, 
J.~Valenti\altaffilmark{6}, 
D.~Barrado\altaffilmark{7},
H.~Bouy\altaffilmark{7},
D.~Ciardi\altaffilmark{8},
M.~Pinsonneault\altaffilmark{9},
K.~Stassun\altaffilmark{10},
G.~Micela\altaffilmark{11},
S.~Aigrain\altaffilmark{12},
F.~Vrba\altaffilmark{13},
G.~Somers\altaffilmark{9},
E.~Gillen\altaffilmark{12,14},
A.~Collier Cameron\altaffilmark{15}}

\altaffiltext{1}{Infrared Science Archive (IRSA), Infrared Processing
and Analysis Center (IPAC), 1200 E.\ California Blvd., California
Institute of Technology, Pasadena, CA 91125, USA; rebull@ipac.caltech.edu}
\altaffiltext{2}{Spitzer Science Center (SSC), Infrared Processing and
Analysis Center (IPAC), 1200 E.\ California Blvd., California
Institute of Technology, Pasadena, CA 9112, USA5}
\altaffiltext{3}{Universit\'e de Grenoble, Institut de Plan\'etologie
et d'Astrophysique de Grenoble (IPAG), F-38000 Grenoble, France;
CNRS, IPAG, F-38000 Grenoble, France}
\altaffiltext{4}{NASA Ames Research Center, Kepler Science Office,
Mountain View, CA 94035, USA}
\altaffiltext{5}{Astronomy Department, California Institute of
Technology, Pasadena, CA 91125, USA}
\altaffiltext{6}{Space Telescope Science Institute, 3700 San Martin Drive,
Baltimore, MD 21218, USA; Center for Astrophysical Sciences, Johns Hopkins University,
3400 North Charles St., Baltimore, MD 21218, USA}
\altaffiltext{7}{Centro de Astrobiolog\'ia, Dpto. de
Astrof\'isica, INTA-CSIC, E-28692, ESAC Campus, Villanueva de
la Ca\~nada, Madrid, Spain}
\altaffiltext{8}{NASA Exoplanet Science Institute (NExScI), Infrared
Processing and Analysis Center (IPAC), 1200 E.\ California Blvd.,
California Institute of Technology, Pasadena, CA 91125, USA}
\altaffiltext{9}{Department of Astronomy, The Ohio State University,
Columbus, OH 43210, USA; Center for Cosmology and Astroparticle
Physics, The Ohio State University, Columbus, OH 43210, USA}
\altaffiltext{10}{Department of Physics and Astronomy, Vanderbilt
University, Nashville, TN 37235, USA; Department of Physics, Fisk
University, Nashville, TN 37208, USA}
\altaffiltext{11}{INAF - Osservatorio Astronomico di Palermo, Piazza
del Parlamento 1, 90134, Palermo, Italy}
\altaffiltext{12}{Department of Physics, University of Oxford, Keble Road, Oxford OX3 9UU, UK}
\altaffiltext{13}{US Naval Observatory, Flagstaff Station, P.O. Box
1149, Flagstaff, AZ 86002, USA}
\altaffiltext{14}{Astrophysics Group, Cavendish
Laboratory, J.J. Thomson Avenue, Cambridge CB3 0HE, UK}
\altaffiltext{15}{School of Physics and Astronomy, University of 
St Andrews, North Haugh, St Andrews, Fife KY16 9SS, UK}

\begin{abstract}  

We use K2 to continue the exploration of the distribution of rotation
periods in Pleiades that we began in Paper I. We have discovered
complicated multi-period behavior in Pleiades stars using these K2
data, and we have grouped them into categories, which are the focal
part of this paper. About 24\% of the sample has multiple, real
frequencies in the periodogram, sometimes manifesting as obvious
beating in the light curves.  Those having complex and/or structured
periodogram peaks, unresolved multiple periods, and resolved close
multiple periods are likely due to spot/spot group evolution and/or
latitudinal differential rotation; these largely compose the slowly
rotating sequence in $P$ vs.~\vmk\ identified in Paper I.  The fast
sequence in $P$ vs.~\vmk\ is dominated by single-period stars; these
are likely to be rotating as solid bodies. Paper III continues the
discussion, speculating about the origin and evolution of the period
distribution in the Pleiades.

\end{abstract}

\section{Introduction}
\label{sec:intro}

Ultra-high precision photometry can reveal a wealth of multi-frequency
behavior, and indeed, one of the original goals of the Kepler mission
was to investigate asteroseismology (\eg, Metcalfe \etal\ 2009) and
stellar cycles, including differential rotation (\eg, Karoff \etal\
2009). Rotation rates and timescales for stellar cycles are thought to
be be a function of age of the star. The original Kepler field,
however, had no clusters younger than 1 Gyr. With the repurposed K2
mission (Howell \etal\ 2014) comes the opportunity to study clusters
of a much wider range of ages, with the potential to constrain
measures of, say, surface differential rotation at much younger ages.

The Pleiades is populous (over 1000 members, \eg, Bouy \etal\ 2015),
relatively young (125 Myr; Stauffer \etal\ 1998a), and nearby (136 pc;
Melis \etal\ 2014). As such, it provides a nearly ideal laboratory for
studying rotation of young stars, including surface differential
rotation and spot or spot group evolution.  These stars are old enough
to have completed all of their pre-main sequence accretion and most of
their pre-main sequence contraction, but they are still young enough
to be rotating quickly enough that there are at least 6-7 complete
rotations over a K2 campaign ($\sim$70 d long), and their spots/spot
groups are large enough to produce brightness variations easily
detectable by K2 ($\gtrsim$0.003 mag).  

Rebull \etal\ (2016), Paper I, presented the K2 Pleiades data and
reduction, membership selection, and our general results.  About 92\%
of the members in our sample have at least one measured spot-modulated
rotation period. The overall relationship between $P$ and \vmk\ 
follows the overall trends found in other Pleiades studies (\eg, Covey
\etal\ 2016, Hartman \etal\ 2010). There is a
slowly rotating sequence for $1.1\lesssim$\vmk$\lesssim 3.7$
(2$\lesssim P \lesssim$11 d), and a primarily rapidly rotating
population for \vmk$\gtrsim 5.0$ (0.1 $\lesssim P \lesssim$ 2 d). 
There is a region in which there seems to be a disorganized
relationship between $P$ and \vmk\ between $3.7\lesssim$\vmk$\lesssim
5.0$ (0.2$\lesssim P \lesssim$ 15 d). 

In the analysis conducted as part of Paper I, we noticed that a
significant fraction of the periodograms, $\sim$30\%, had additional,
significant peaks in the periodogram. In these cases, the false alarm
probability (FAP) for the additional peaks was just as low as that for
the main peak, often calculated to be exactly 0. The K2 light curves
(LCs) are so exquisite that when subsidiary peaks appear as
significant, it is worth investigating, which is the motivation for
this paper. 

In this paper, we focus on the diversity of multi-frequency
periodograms and LCs we find in these K2 Pleiades data.
Section~\ref{sec:obs} summarizes the key points of the observations
and data reduction from Paper I.  Section~\ref{sec:multifreq}
describes the several categories of LCs and periodogram structures
that we found in the K2 data.  The simplest division of the LCs and
periodograms is into single- and multi-period stars; the distributions
of these broad categories are discussed in
Section~\ref{sec:singlemulti}.  However, we have many more than just
two categories of LCs and periodogram structures;
Section~\ref{sec:alltherest} looks at where these various categories
of objects fall. Section~\ref{sec:deltaP} looks at the distribution of
period differences. Section~\ref{sec:concl} summarizes our main
results.  This is the second of three papers focused on rotation
periods in the Pleiades. Stauffer \etal\ (2016), Paper III, continues
the discussion of these Pleiades results, focusing on the physical
origins of the $P$ distribution.

\section{Observations and Methods}
\label{sec:obs}

The observations and methods are discussed in detail in Paper
I. Here, we simply summarize the main points:
\begin{itemize}
\item Members of the Pleiades were observed in K2 campaign 4, which lasted
for 72d.  All of the stars in this sample were observed in the
long-cadence ($\sim$30 min exposure) mode. 
\item We looked for periods using the 
Lomb-Scargle (LS; Scargle 1982) approach.
\item For stars of the mass range considered here, the periods that we
measure are, by and large, star spot-modulated rotation periods. 
Spot modulation is the simplest explanation for sinusoidal (or
sinusoidal-like) variations where there are changes over an entire
orbital phase. 
\item We assembled a catalog of literature data for our targets. The
most important values obtained from this search are \vmk\ (measured or
inferred; see Paper I) and membership (see Paper I). Cross-IDs between
the EPIC number, RA/Dec, and common literature names are in Paper I.
\item Out of the 1020 LCs of candidate Pleiades members from which we
started,there are 775 high-confidence members, with 51 more lower-confidence
members (for a total of 826 members). Table~\ref{tab:lccounts} includes
these numbers.
\item Out of those 775 (best members), 716 (92.4\%) have at least one
measured period that we believe in the overwhelming majority of cases to be
a rotation period and due to star spots. Including the lower confidence
members, 759/826 (91.9\%) have at least one measured period. 
\item The period distribution is strongly peaked at $P<$1d with
typical amplitudes of $\sim$0.03 mag.
\end{itemize}

Table~\ref{tab:bigperiods} includes the most relevant supporting data
from Paper I, plus, for each star, its LC categories as introduced in
this paper.

As stated above, what drove us to investigate the multi-period
behavior was the presence of additional, significant (low FAP) peaks
in the periodogram. To first order, we took the periods in order of
peak strength (and thus significance); all other things being equal,
the highest peak is what we took to be the primary period, $P_1$, the
second highest peak is what we took to be the secondary peak, $P_2$,
etc. However, in some cases, we had to overrule that ordering of the
periods (see, \eg, \S\ref{sec:mws}).

\begin{deluxetable}{cl}
\tabletypesize{\scriptsize}
\floattable
\tablecaption{Contents of Table: Periods, Supporting Data, and Light Curve
Categories for Periodic Pleiades\label{tab:bigperiods}}
\tablewidth{0pt}
\tablehead{\colhead{Label} & \colhead{Contents}}
\startdata
EPIC & Number in the Ecliptic Plane Input Catalog (EPIC) for K2\\
coord & Right ascension and declination (J2000) for target \\
Vmag & V magnitude (in Vega mags), if observed\\
Kmag & \ks\ magnitude (in Vega mags), if observed\\
vmk0 & \vmk\ -- dereddened $V-K_s$, directly observed (if $V$ and \ks\
exist) or inferred (see text)\\
P1 & Primary period, in days (taken to be rotation period)\\
P2 & Secondary period, in days\\
P3 & Tertiary period, in days\\
P4 & Quaternary period, in days\\
ampl & Amplitude, in magnitudes, of the 10th to the 90th percentile\\
memb & membership indicator: Best, OK, or NM \\
single/multi-P & indicator of whether single or multi-period star\\
dd &  indicator of whether or not it is a double-dip LC \\
ddmoving & indicator of whether or not it is a moving double-dip LC\\
shch & indicator of whether or not it is a shape changer\\
beat & indicator of whether or not the full LC has beating visible\\
cpeak & indicator of whether or not the power spectrum has a complex,
structured peak and/or has a wide peak\\
resclose & indicator of whether or not there are resolved close
periods in the power spectrum\\
resdist & indicator of whether or not there are resolved distant
periods in the power spectrum\\
pulsator & indicator of whether or not the power spectrum and period
suggest that this is a $\delta$ Scuti pulsator\\
cloud & indicator of whether or not the phased light curve has narrow,
angular dips\\
\enddata
\end{deluxetable}

\clearpage

\begin{deluxetable}{p{2cm}cccp{6.5cm}p{4cm}}
\tabletypesize{\scriptsize}
\floattable
\tablecaption{Star/Light Curve/Periodogram Categories\tablenotemark{a}
\label{tab:lccounts}}
\tablewidth{0pt}
\tablehead{\colhead{Name} & \colhead{Number} & \colhead{Frac.} & \colhead{Frac.\ of} & \colhead{Empirical
 properties} & \colhead{Possible Physical Interpretation}\\
 & &  \colhead{of
sample\tablenotemark{b}} & \colhead{periodic sample\tablenotemark{c}}  }
\startdata
Initial sample (see Paper I)	&	1020	& \ldots & \ldots &	All K2 LCs of candidate Pleiades members	& \ldots \\ 
Best members (see Paper I)	& 775 & \ldots & \ldots&	Highest-confidence (our determination)
Pleiades members with K2 light curves, and neither too bright nor too
faint (6$<$\ks$<$14.5)   	&\ldots \\ 
OK members (see Paper I)	& 51 & \ldots & \ldots&	Lower-confidence (our determination) Pleiades
members with K2 light curves, and neither too bright nor too faint
(6$<$\ks$<$14.5)   	&\ldots \\ 
The sample, a.k.a. the sample of members	& 826 & 1.00 & \ldots&	The set of all
high-confidence (`best') plus lower-confidence (`ok') members that
are  neither too bright nor too faint (6$<$\ks$<$14.5)	& \ldots \\ 
The periodic sample &	759	& 0.92 & 1.00 &	The subset of all
high-confidence (`best') plus lower-confidence (`ok') members that
are  neither too bright nor too faint (6$<$\ks$<$14.5) and are found
to be periodic by us in these K2 data 	& 
\ldots	\\ 
\hline					
Single frequency	&	559 & 0.68 & 0.74	&	The star has just one
detectable period. Could be sinusoid or other shapes, but there is
just one detectable period. 	&	single spot/spot group
rotating into and out of view \\ 
Double-dipped (or double-humped) &	107 & 0.13 & 0.14 &	There is a
double-dipped (or double-humped) structure in the phased LC, {\em and}
the power spectrum has a substantial amount of power in the harmonic
($P/2$) compared to the main $P$. 	&	two spot/spot groups,
well separated in longitude, rotating into and out of view \\ 
Moving double-dip	& 31 & 0.038 & 0.041 & 	{\em (Effectively a subcategory of both
double-dip and shape changers.)} There is a double-dipped (or
double-humped) structure in the phased LC, {\em and} often the power
spectrum has a substantial amount of power in the harmonic ($P/2$)
compared to the main $P$, {\em and} one can see by eye that one
structure in the LC is moving with respect to the other structure over
the campaign.	& spot/spot group evolution and/or latitudinal
differential rotation	\\
Shape-changer	& 	114 & 0.14 & 0.15 &  Shape of the light curve changes
over the campaign, but not enough such that a separate period can be
derived. Usually, these are single-period stars.	& 
latitudinal differential rotation and/or spot/spot group evolution\\ 
Orbiting clouds?	& 	5 & 0.0061 & 0.0066 & 	The phased light curves show
shallow, angular dips which cover a relatively small portion of the
total phase. (NB: one more is formally a NM, for a total of 6) &
orbiting clouds or debris?? \\ 
Multi-frequency & 	200 & 0.24 & 0.26 & 	More
than one frequency is measured, or apparent by eye (in the periodogram
or the LC itself) even if specific
value cannot be determined.	& 	spot/spot group evolution
and/or latitudinal differential rotation and/or binarity \\
Beater	& 	135 & 0.16 & 0.18 & 	Light curve appears to have beating
signatures by eye (\eg, changing envelope over the campaign).
&	spot/spot group evolution and/or latitudinal differential
rotation	\\ 
Complex peak & 	89 & 0.11 & 0.12 & 	Peak in power
spectrum is structured (multi-component) or wider than expected for
that period. & 	latitudinal differential rotation and/or spot/spot
group evolution 	\\ 
Resolved multi-period, close & 126 & 0.15 & 0.17 & 	Well-resolved, very narrow peaks in the power
spectrum and both periods are real. The periods are very close
together by the $\Delta P/P$ metric.	& latitudinal differential rotation 
and/or spot/spot group evolution in most cases;
binarity in other cases \\ 
Resolved multi-period, distant	& 37 & 0.045 & 0.049 & Well-resolved, very narrow peaks in the power spectrum and
both periods are real. The periods are {\em NOT} very close together
by the $\Delta P/P$ metric.	&  binarity	\\ 
Pulsator	& 8 &0.0097 & 0.011 &	``Forest'' of
short-period peaks in the periodogram.	&	Pulsation ($\delta$
Scuti)\\ 
\enddata
\tablenotetext{a}{The first half of this table includes numbers discussed
in detail in Paper I. The second half of this table is a high-level summary
of the categories discussed in detail in Sec.~\ref{sec:multifreq}. Please
see the text for a much longer discussion of properties and
interpretation.  }
\tablenotetext{b}{Fraction out of `the sample', \eg, 826.  Stars can
belong to more than one category, so the fractions do not add to 1.}
\tablenotetext{c}{Fraction out of `the periodic sample', \eg, 759. 
Stars can belong to more than one category, so the fractions do not
add to 1.}
\end{deluxetable}

\clearpage

\section{Empirical Structures in the Light Curves and Power Spectra}
\label{sec:multifreq}
\label{sec:types}

While inspecting the power spectra of the ensemble of K2 LCs, about
60\% of the ensemble had properties as found in Fig.~2 of Paper I,
where there was just one sinusoidal periodic signal, and it was
unambiguously periodic (very low FAP), with just one period. 
However, we noted that about 30\% of the sample potentially had more
than one significant period in the power spectrum, where the FAP was
just as low for these additional peaks. We phased the LCs at these
additional peaks, and in many cases, the phased LC is just as
convincing as the LC phased at the main peak. 

We noticed patterns in the power spectra (and/or the original and
phased LCs), which enabled us to group these multi-period objects into
categories, which we discuss in this section, and summarize in
Table~\ref{tab:lccounts}. Note that objects can simultaneously belong
to more than one of these categories.  We conclude this section by
discussing the ensemble incidence rate of multi-period LCs.  For all
of the discussion here, `the sample' is members of the right
brightness range, as described in Paper I and
Table~\ref{tab:lccounts}.

For those stars where we can distinguish multiple plausible
frequencies, we have included in Table~\ref{tab:bigperiods} the
secondary (and tertiary or even quaternary in some cases) periods. The
first period that is listed is the strongest period and/or the one we
believe to be the rotation rate of the star. Some of these stars are
known or inferred binaries; in these cases, the period should most
often be that of the primary. If they are unresolved multiples, we
cannot assign directly measured \ks\ or \vmk\ to the companion, so
these companions do not appear as separate stars in our analysis (here
or in Paper I or III). 

Finally, for completeness, we note that in a few cases, it was
difficult to determine if a subsidiary peak in the peridogram was a
harmonic (usually half the true $P$) or an independent period.  Our
primary discriminent was by eye -- if the LC phased to a subsidiary
peak had characteristics of a harmonic (\eg, a wide distribution of
fluxes at most phases and a `criss-cross' appearance of the phased
LC), then we took it to be a harmonic and did not include it as an
independent period derived from our data.  For most of these cases
where the by-eye assessment identified the subsidiary peak as a
harmonic, the ratio of the true $P$ to the harmonic was very close to
2.  Based on that, even if the phased LC did not obviously appear to
be a harmonic, but the ratio of the periods was within 3\% of 2 (1.97
$< P1/P2<$2.03), then we took the periods to be, in fact, harmonics,
and did not retain the shorter $P$ as an independent, second period. 


\subsection{Double-dipped LCs}
\label{sec:mws}

\begin{figure}[ht]
\epsscale{1.0}
\plotone{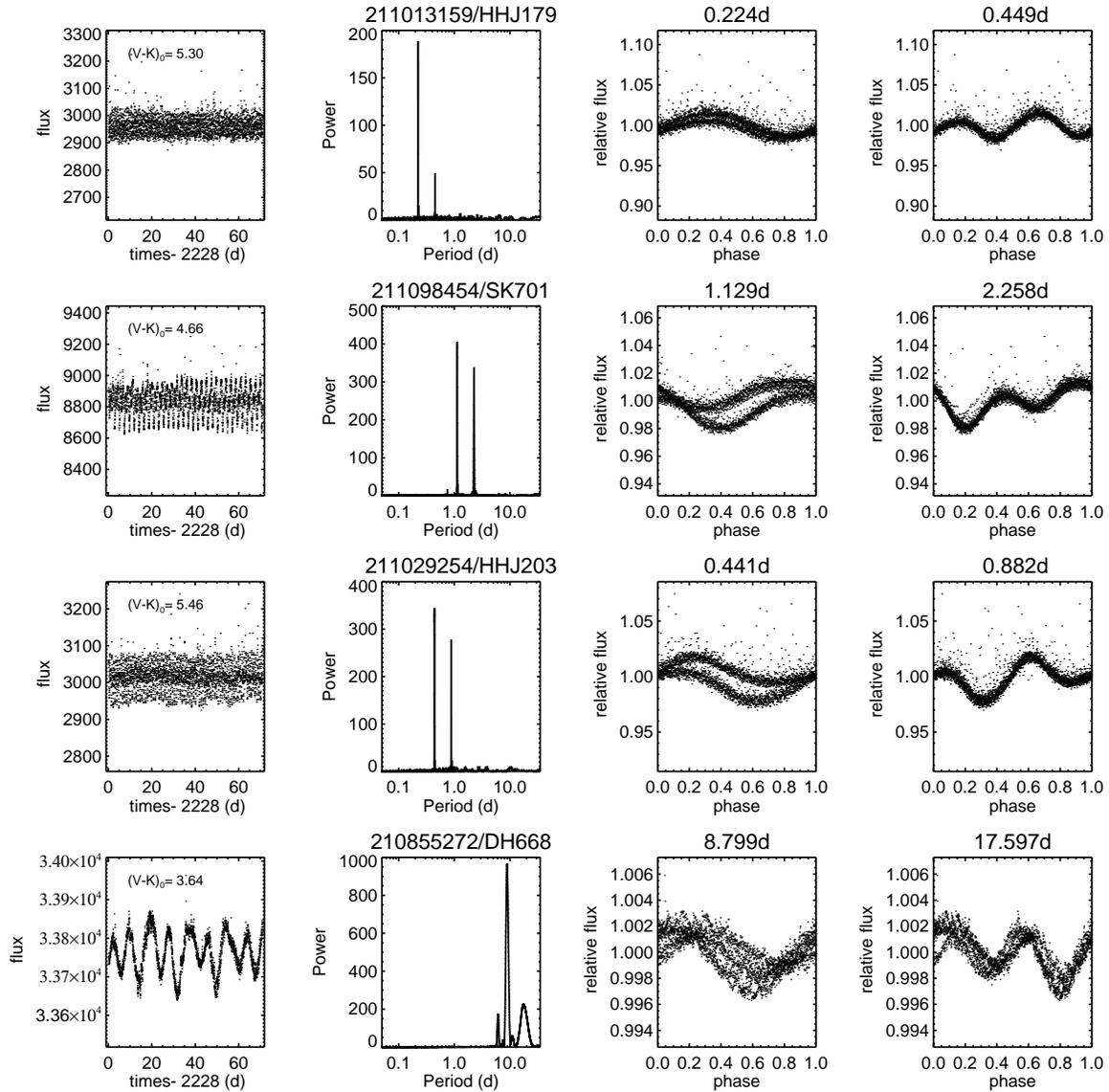}
\caption{Four examples of the double-dipped category, where the 
double-dipped structure in the LC
`fools' the periodogram such that the maximum peak in the periodogram
is half the true period. First column: full LC (with \vmk\ for star as
indicated); second column: LS
periodogram; 3rd column: phased LC with $P$ corresponding to the
maximum peak in the periodogram; 4th column: phased LC, with best
period (in days) as indicated. Rows, in order: EPIC 211013159/HHJ179, 
211098454/SK701,  211029254/HHJ203,   and 210855272/DH668. These are
representatives from a range of brightnesses and periods. In these
cases, the double-dipped structure redistributes power from $P$ into
$P/2$ because the dips are of comparable depth. These LCs most likely
result from stars with two spots or spot groups on well-separated
longitudes. (NB: 210855272/DH668 is discussed in Paper I, because the
longer period identified here makes it an outlier in the $P$ vs.~\vmk\
plot, but the shorter period would move it `into line' with other
stars of its \vmk. However, we believe the longer period is correct
for this star.)  }
\label{fig:mw-still}
\end{figure}

For sinsusoidal, single-period LCs, as seen in Fig.~2 of Paper I,
if there is a secondary peak in the power spectrum, it is very weak
relative to the main peak. However, many of the LCs  have
double-dipped (or double-humped) structure in their phased
LCs\footnote{McQuillen \etal\ (2013) use this same terminology for
similar LCs.}.  When the two dips are of comparable depth and fraction
of the orbital phase, there is enough power in the $P/2$ harmonic that
it appears with comparable strength in the periodogram. Taken at face
value, then, such a power spectrum makes the star appear to have (at
least) two significant periods; see Figure~\ref{fig:mw-still}.
Sometimes the $P/2$ peak is of comparable strength to the $P$ peak,
and sometimes, the $P/2$ harmonic is actually the dominant peak
(perhaps by a considerable margin) in the periodogram. (Recall that
the LS approach is fitting sinusoids, so the relative strength of the
peaks is related to the strength of the single sinusoidal nature of
the periodicity.) For these cases, visual inspection reveals that the
true rotation period is actually $P$ and not $P/2$, as shown in
Fig.~\ref{fig:mw-still}.  The peaks are typically very narrow,
suggesting a limited range of frequencies found around the period (and
$P/2$) during the campaign.   About 14\% of the sample (and 15\% of
the periodic sample) have this kind of power spectrum and phased LC
shape. 

\begin{figure}[ht]
\epsscale{1.0}
\plotone{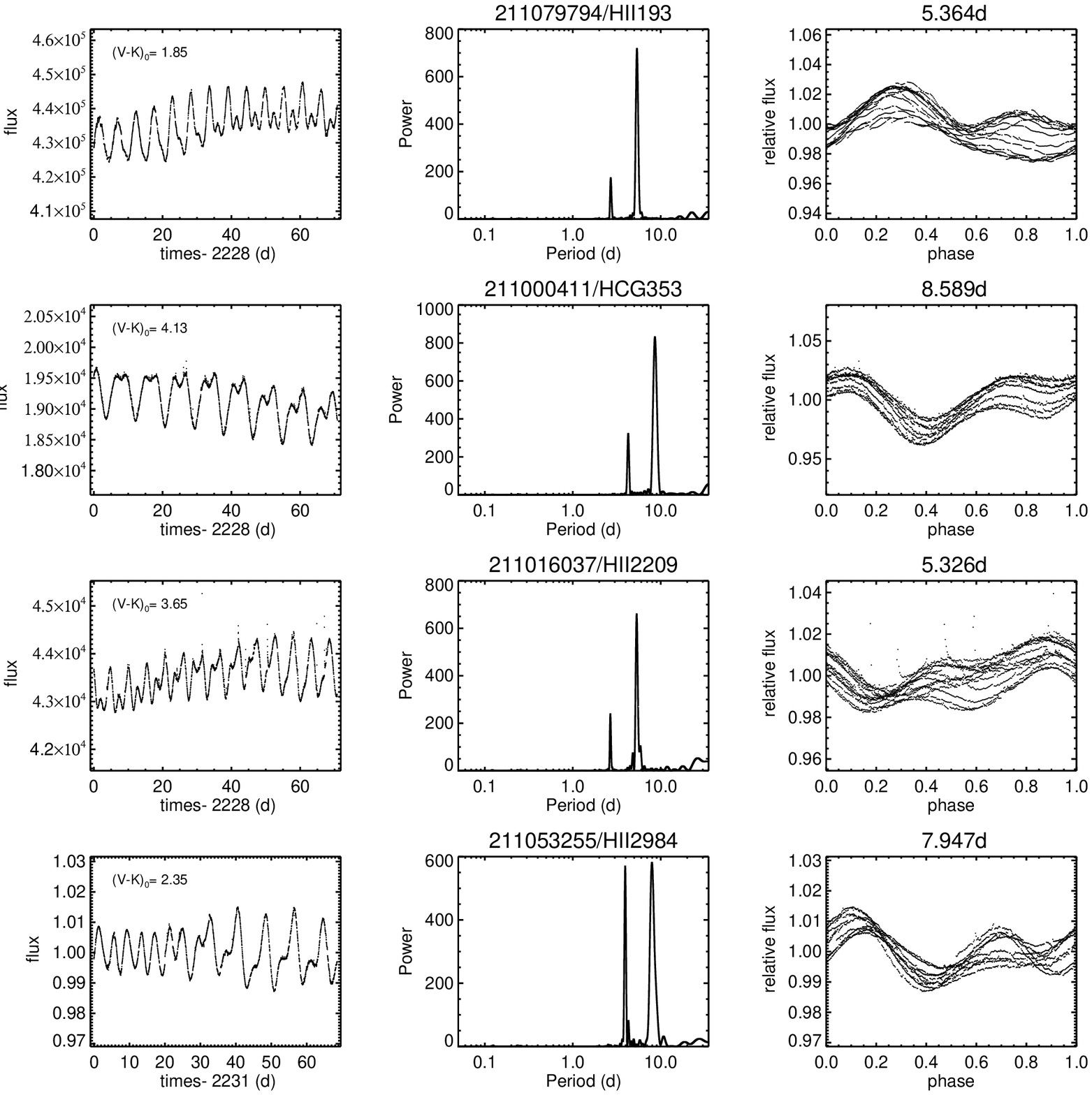}
\caption{Four examples of the moving double-dipped category, 
where there is a double-dipped structure in
the LC, as in prior figure, but there are visible changes in the
relative locations of the  structure over the campaign. First column:
full LC (with \vmk\ for star as
indicated); second column: LS periodogram; 3rd column: phased LC, with
best period (in days) as indicated.  Rows, in order:
211079794/HII193,  211000411/HCG353,  211016037/HII2209, and
211053255/HII2984. These LCs most likely result from stars with two
spots or spot groups at different latitudes with differential rotation
and/or spot evolution. Also see \S\ref{sec:shch}.}
\label{fig:mw-moving}
\end{figure}

We identified the double-dip LC types by eye, but we considered both
the properties of the phased LC shape and the structures of the
periodogram. To be in the double-dip category, there has to be 
double-dip (or -hump) structures in the phased LC as well as
substantial power in the nominally subsidiary peak $P_{\rm true}/2$.
(In other words, the periodogram has to look like it has 2 significant
periods, but in reality only one of the pair is a viable
period.\footnote{This feature, where the periodogram looks like it has
two significant periods but in reality has only one real period,
accounts for the apparent discrepancies in some of the sample
fractions given. At the end of the analysis: (a) 59\%  have one and
only one period and a $\sim$sinusoidal LC. (b) 68\% have one and only
one period, consisting of both the $\sim$sinusoidal LC and those double-dip
LCs that look like they have 2 significant periods but in reality have
only one. (c) When approaching the analysis of the LCs with {\em
potentially} more than one significant peak in the periodogram,
therefore, we include all the stars that in the end legitimately have
more than one period, plus the double-dip stars that only appear to
have more than one period (noting that some double-dip stars do
legitimately have more than one real period). Therefore, 33\% of the
sample have {\em potentially} more than one significant peak in the
periodogram, which then decreases to 24\% actually having multiple
frequencies in the end. (8\% are not periodic at all.)}) Therefore, we do not
include in this category those LCs that have multiple dips/humps in
the phased LC where the ratio of the peak power values for the correct
$P$ and the first subsidiary peak is large.  A phased LC with two (or
more) dips of very different shape or relative duration will result in
the power spectrum correctly identifying the right period, with little
power in the harmonics. (For examples of LCs where there are multiple
dips but nearly all the power is in the correct $P$, see the
single-period examples in Fig.~\ref{fig:batwings} below.)  For all the
single-period, $\sim$sinusoidal LCs in our set, the median ratio of
the power of the main peak (in the periodogram) to the second peak
power is $\sim$14. For the double-dip LCs, the median of this power
ratio is $\sim$2.  

In some cases, inspection of the LC shows a double-dip structure over
only part of the K2 campaign, or there are other slow changes over the
K2 campaign such that the double-dip structure may be transient or the
location of one of the minima changes slowly with respect to the
maxima over the campaign. These changes are not large enough to, on
their own, generate enough power in a secondary period, so no
secondary period can be identified; these changes also mean that the
power may not be redistributed into a subsidiary peak to the same
degree as the rest of this category for a periodogram calculated over
the entire K2 campaign. These so-called ``moving double-dip'' stars
are identified by eye and are a subcategory of the overall double-dip
category; examples are given in Fig.~\ref{fig:mw-moving}.  (These
moving double dip stars are also `shape changers'; see
Sec.~\ref{sec:shch} below.)

Most of the stars in this category have only one real period. However,
some of these double-dip LCs have at least one additional periodic
signal, which may or may not also have the same double-dip effect that
rearranges the power distribution.  (For an example where two periods
are found that are both double-dip, see 211054634/HHJ291 in
Fig.~\ref{fig:resolved} below.)

For completeness, we add that there are a few borderline cases (such
as 211048126/HII1321) where the power spectrum has some signatures of
this sort of effect (harmonic at $2P$ rather than $P/2$, and/or strong
harmonic at $P/2$), but the phased LC is not convincing. Those have
not been included in this category.

These double-dip light curve structures are most likely to be caused
by spots well-separated in longitude on the star, as in Davenport
\etal\ (2015).   The Fourier transform of the light curve of a low
latitude spot viewed equator-on on a limb-darkened star has
significant power in its first harmonic because there is a flat
plateau when the spot is behind the star. The resulting light curve is
well approximated by a combination of the fundamental and its first
harmonic. For this reason, starspot distributions of arbitrary
complexity always give either single-humped or double-humped light
curves, the exact morphology depending on the degree of departure from
axisymmetry. If there are two concentrations of spots on opposite
hemispheres, the fundamental cancels but the first harmonics from both
groups reinforce each other, giving an apparently stronger signal in
the first harmonic. (Also see Russell 1906 for a discussion of similar
phenomena in the context of asteroid shapes.) 

There is no preferential \vmk\ color for double-dip objects on the
whole (see discussion in \S\ref{sec:alltherest} below); they are found
at all colors for which we have K2 spot-modulated light curves, as
would be expected for spot periods. However, there are strong
correlations between the stability of the pattern and \vmk; the moving
double-dip are all earlier types, with 1$\lesssim$\vmk$\lesssim$4.6
(FGK and early M). Almost all the stationary double-dip objects have
\vmk$\gtrsim$4; there are a few with 3$\lesssim$\vmk$\lesssim$4. 

Those that show movement of one dip/hump with respect to the other over
the K2 campaign could be explained by spot/spot group formation or
migration with strong latitudinal differential rotation. Those that do
not show such changes, therefore, are more likely to be rotating as
solid bodies, or at least have weaker latitudinal differential
rotation and/or more spot stability. Although there is a huge range of
$P$ for moving and stationary double-dip objects, the moving
double-dip objects are on the whole rotating slower than the
stationary double-dip objects. Faster rotation should yield a more
dipolar magnetic field and more stable spots/spot groups; see
discussion in Paper III. This is consistent with the results from
Zeeman-Doppler imaging of M dwarfs (Morin \etal\ 2008, 2010).


\subsection{Beating LCs}
\label{sec:beating}

About 16\% of the sample (18\% of the periodic sample) have obvious
signatures of beating in the LC, such as changes in the envelope over
the campaign; see Figure~\ref{fig:complexpeak}.  Most of those have 
some combination of obvious beating in the LC and multiple
frequencies in the periodogram (but are not the double-dip phenomenon
above). For these, phasing to the maximum peak in the periodogram does
not show a very well-phased LC, because of the structure provided by
the subsidiary peak(s). Phasing the LC at the subsidiary peak(s) shows
similar coherence (or lack thereof) for the same reason, even when the
false alarm probability (FAP) is low.

We identified LCs with these kinds of properties initially by
eye based on structure in the LC or in the power spectrum. We
then performed an additional LS periodogram analysis with much finer
frequency sampling than our initial approach, but only between $P$/3
to 3$P$, for the period with the maximum power.  We noticed primarily
two kinds of periodogram structure. One kind has a measurably wide
primary peak, and/or an asymmetric primary peak. The other kind has
peaks that are very narrow, and the main power peak breaks into two
(or sometimes more) resolved components. Often a secondary peak that
appeared at first glance to be a harmonic is not, in fact, a true harmonic
of the main peak. Sometimes the multi-peaked or asymmetric structure
seen in the main peak can also be found in its harmonic (see below and
Fig.~\ref{fig:complexpeak}). 

We discuss the complex power spectrum peaks and those with resolved
multiple frequencies in the next two subsections, respectively. We
note, however, that in a few cases, the LC is seen to have a beating
structure, but the components are not resolved in the power spectrum;
that is, the peak is still thin, and no other significant peaks are
noted. In a few cases, it seems that the {\em frequency} is not
changing, but the {\em amplitude} of one or both of the signals is
changing.  In the remainder of the cases, it seems that the periods
are so close together that the 72-day campaign does not provide enough
cycles to distinguish the periods.

When the LC looks `beating', and/or there are complex power spectrum
peaks, there are 3 possible interpretations: (a) spots or spot groups
evolving; (b) latitudinal differential rotation from spots/spot groups
at more than one latitude during the K2 campaign; (c) spot modulated
periods from two (or even three) unresolved stars in the K2 aperture.
The majority of the stars identified as beaters are the earlier types,
\vmk$\lesssim$4, which is where strong latitudinal differential
rotation is expected. However, hotter stars have shorter-lived spots
in general, so spot/spot group evolution may also be important.


\subsection{Complex, Structured Peaks and Unresolved Multi-Period
Peaks}
\label{sec:complexpeak}

\begin{figure}[ht]
\epsscale{1.0}
\plotone{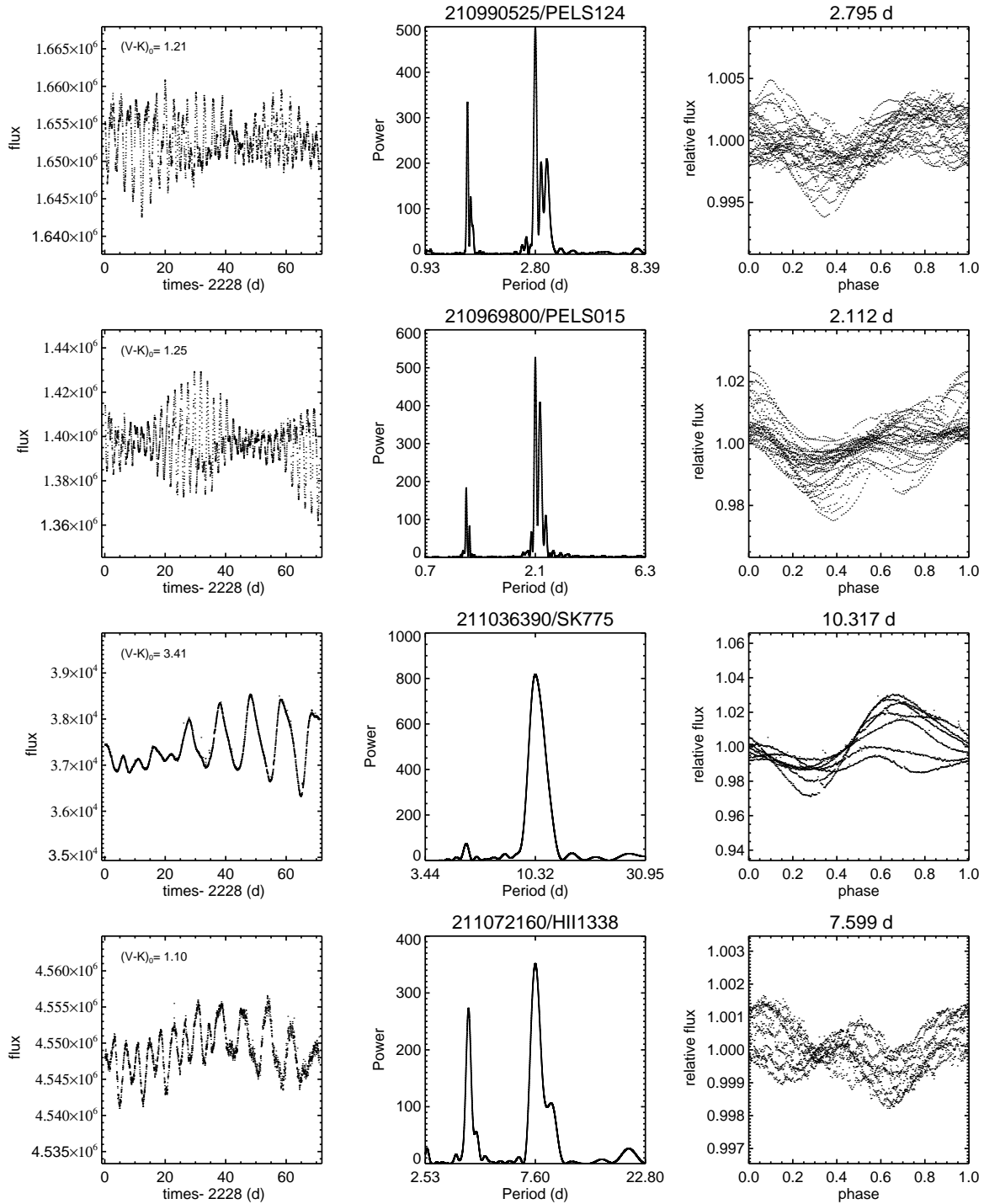}
\caption{Four examples of the beater and complex peak categories.
Beating is seen in the LC, and for these examples, the peak in the
periodogram is complex, broad, and/or structured. Left column: full
LC (with \vmk\ for star as
indicated); middle column: periodogram; right column: phased LC for maximum
peak. Rows, in order (with the periods corresponding to significant
peaks in parentheses): 210990525/PELS124 (2.795, 1.421, 3.330), 
210969800/PELS015 (2.112, 2.212),  211036390/SK775 (10.317), and 
211072160/HII1338 (7.599, 8.899). Note that the power spectrum shows
peaks of measurable width and/or multiple peaks. Sometimes the
multi-peaked structure can be found both in the main peak and its
harmonic. These LCs are likely a result of latitudinal differential
rotation and/or spot evolution.}
\label{fig:complexpeak}
\end{figure}

To find objects where the peak(s) in the periodogram are distinctly
broad and/or structured, we took multiple approaches; examples are
shown in Fig.~\ref{fig:complexpeak}. About 11\% of the sample (12\% of
the periodic sample) falls into this category of complex, structured
peaks and unresolved multi-period.

First, for all of these objects, we computed a power spectrum with
finer frequency resolution, as discussed above.  If the maximum peak
is clearly asymmetric (such as seen in the first and last rows of 
Fig.~\ref{fig:complexpeak}), then we added the star to this category
on the basis of having ``complex, structured peaks.'' If the higher
frequency resolution power spectrum incompletely resolves the peak
into pieces (as seen in the second row of Fig.~\ref{fig:complexpeak},
where the wings of the peaks run into each other at power levels
$>$0), then we added the star into this category on the basis of
having a ``complex peak.''

Finally, to assess whether or not the peak is broader than expected
for its period, we took the full-width-at-half-max (FWHM) of the
actual peak ({\em not} the FWHM of, say, a Gaussian fit to the peak,
but the true FWHM of the peak), and plotted that as a function of
period. For most of the sample, this relationship falls on a line when
plotted in log-log space\footnote{The expected peak width (where $f$
is frequency) is $df \sim 1/T$ (where $T$ is the total length of the
campaign), which is independent of period, and consequently $dP = df 
\times P^2$.}. We fit the line, subtracted it from the distribution,
and looked for things that deviated significantly from 0. This is not
a good approach for identifying, say, objects like the first and last
rows in Fig.~\ref{fig:complexpeak}, because the additional structures
seen near the base of the peaks do not affect the half maximum.
However, the object in the third row of Fig.~\ref{fig:complexpeak} has
a peak much broader than expected for stars with single periods near
10.3d. This is the final type of object added to this ``complex peak''
category -- those with peaks wider than expected for that period. In
these cases, the period (and/or amplitude) is changing over the campaign,
and/or the multiple periods are so close together that the power
spectrum cannot resolve them.

These kinds of power spectra are very suggestive of latitudinal
differential rotation and/or spot evolution. Periods that are very
close together and/or changing over the K2 campaign could easily be
explained if we are seeing spots at different latitudes in a
differentially rotating star, either spots that persist over the K2
campaign or spots that appear, disappear, or otherwise change over the
K2 campaign. The majority of the stars identified as having complex
peaks are the earlier types, \vmk$\lesssim$4, which is where strong
latitudinal differential rotation is expected, but also where spots
may most rapidly evolve. Stars where the complex peak structure of the
primary peak is repeated in the first harmonic may be more likely to
be differential rotation than spot evolution.  On the other hand,
complex peaks close to the rotation period can also arise when active
regions appear and decay at random longitudes, giving brief periods of
coherent modulation with phase shifts between them. The sine function
used in the periodogram interferes constructively with such
widely-separated but phase-shifted modulations at a frequency slightly
different from the true rotation frequency. For a finite data train
lasting a few active-region lifetimes, this splits the fundamental
into several shifted peaks, even in the absence of differential
rotation.\footnote{There is a close analogy with the collisional
spectral-line broadening that results from phase changes induced by
atomic collisions in a gas.} We (ACC and SA) have successfully
simulated these light curves as a Gaussian process with a covariance
function comprising a single modulation period and a spot lifetime of
a few rotations. The shapes of the resulting light curves resemble
these LCs closely, and the pattern of splitting of their periodogram
peaks is very similar.

\subsection{Resolved Multi-Period}
\label{sec:resolvedmultifreq}

\begin{figure}[ht]
\epsscale{1.0}
\plotone{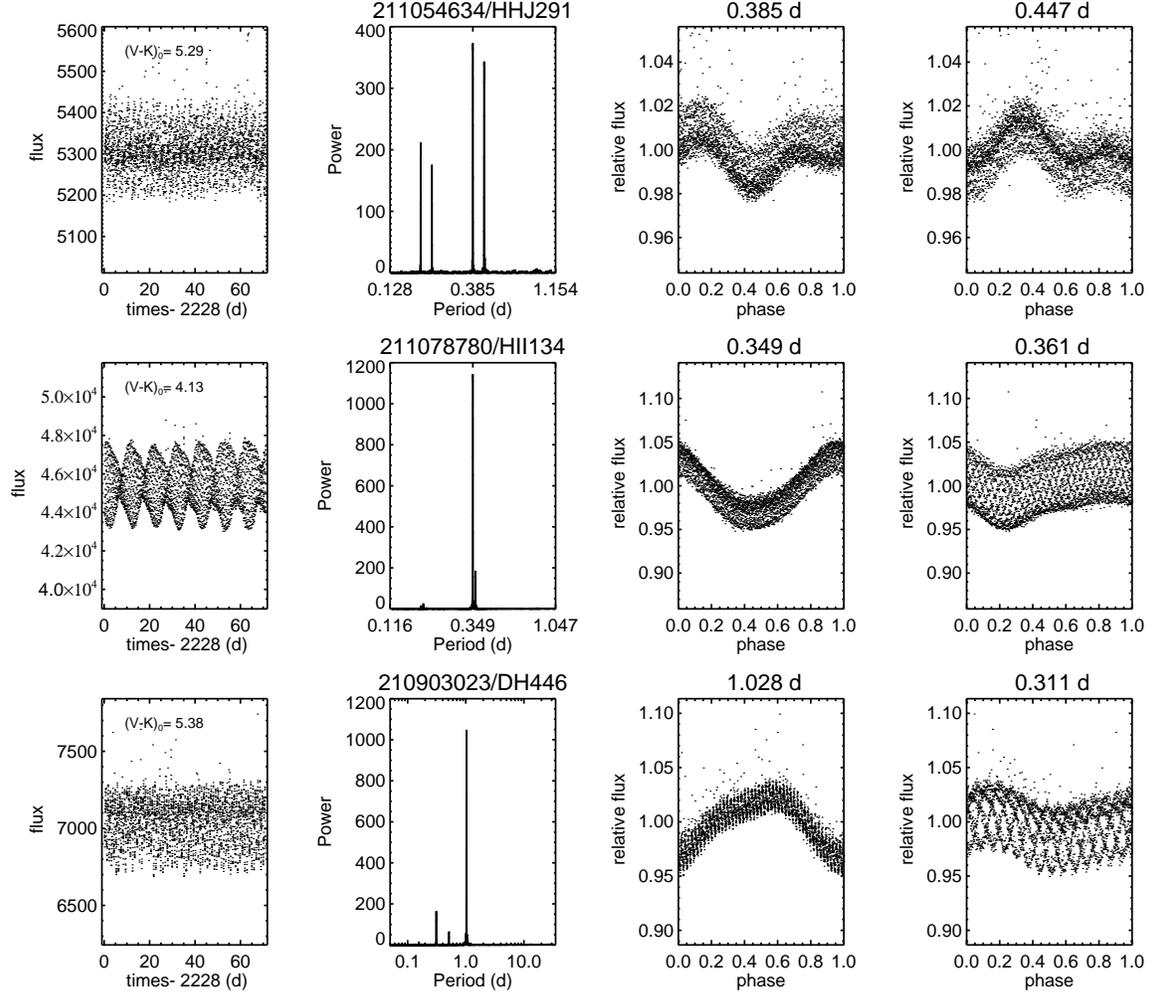}
\caption{Three examples of the resolved peak categories, LCs where
there are 2 distinct, well-resolved frequencies. First column: full LC
(with \vmk\ for star as indicated); 2nd column: periodogram; 3rd
column: phased LC for primary peak; last column: phased LC for
secondary peak. 
Rows, in order (with the periods
corresponding to significant peaks in parentheses):  
211054634/HHJ291 (0.385, 0.447; $\Delta P/P_1$=0.162), 
211078780/HII134 (0.349, 0.361; $\Delta P/P_1$=0.035), 
210903023/DH446  (1.028, 0.312; $\Delta P/P_1$=0.698). 
Note that the power spectrum shows multiple, resolved peaks. Following
the $\Delta P/P_1$ metric divisions, the first and second have
resolved close periods (the two peaks are very close together in the
periodogram), and the third has resolved distant peaks.  All three of
these have companions that can be seen in POSS or Robo-AO data, so
these LCs are likely to result from binaries, with one period per
star. }
\label{fig:resolved}
\end{figure}

In some cases, there are well-resolved, very narrow peaks in the power
spectrum and both periods are real; see Figure~\ref{fig:resolved}. The
periods need not be very close together and they are not harmonics,
though sometimes the harmonics appear (first row of
Fig.~\ref{fig:resolved}).  About 20\% of the sample (22\% of the
periodic sample) fall into this category of resolved periods. 

We wished to differentiate situations as in the second row of
Fig.~\ref{fig:resolved}, where the peaks are very close together, and
situations as in the last row of Fig.~\ref{fig:resolved}, where the
peaks are substantially farther away from each other. For every
situation in which we had at least two viable periods for a star, we
took the closest two periods out of those detected, subtracted the
smaller from the larger, and divided by the primary peak. The primary
peak is the strongest in the power spectrum, and what we take to be
the rotation period of the primary star in the system, but may or may
not be one of the two closest peaks. If that metric, $\Delta P/P_1$,
is less than 0.45, then we take it to be resolved close peaks; if that
metric is greater than 0.45, then we take it to be resolved distant
peaks. This dividing line of $\Delta P/P_1$=0.45 was determined by
inspection of the light curve and periodogram properties, as well as
the morphology of the $\Delta P/P_1$ distribution; see much more
discussion in Sec.~\ref{sec:deltaP} below.

Of the $\sim$20\% of the sample that falls in this broader category of
resolved peaks, 73\% are tagged close, and 30\% are tagged distant.
Note that some objects have both close and distant peaks. Both
resolved close and resolved distant periods are found at all \vmk; the
distant periods are equally likely at all \vmk, but the close periods
are slightly more frequent (as a fraction of the sample) for
\vmk$\lesssim$3.7.

Particularly for the stars with \vmk$\lesssim$3.7 and having resolved close
periods, we suspect that these are another manifestation of
latitudinal differential rotation and/or spot evolution.  For the resolved
distant periods, we suspect that they are binaries, where the two
different periods are the different rotation rates of the components
of the binary. For $\sim$60\% of the stars in this category, there is
at least some evidence that it is a binary or there is more than one
star in the K2 aperture; for the rest, there is no information
suggesting binarity, but in many cases, this may reflect simply a lack
of data.

\subsection{Shape Changers}
\label{sec:shch}

There is a category of LC where the power spectrum suggests that there
is just one period derived from the data, but there are visible,
substantial changes in the shape of the light curve over the campaign.
These so-called shape changers are $\sim$14\% of the sample
($\sim$15\% of the periodic sample). 

\begin{figure}[ht]
\epsscale{0.8}
\plotone{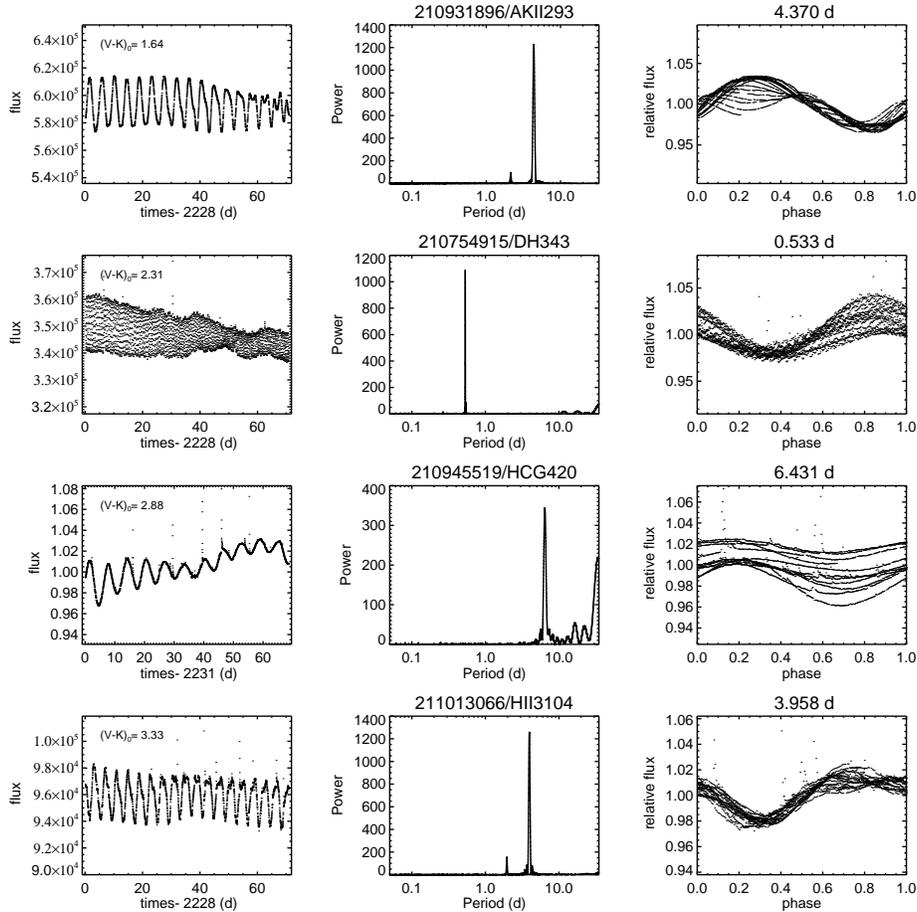}
\caption{Four examples of the shape changer category, LCs where there
are changes in the shape of the LC over the K2 campaign. Left column:
full LC (with \vmk\ for star as
indicated); middle column: periodogram; right column: phased LC for
maximum peak. Rows, in order:  210931896/AKII293, 210754915/DH343,
210945519/HCG420, 211013066/HII3104. Note that the power spectrum
shows a single peak, but the shape changes over the campaign. Assuming
that these shapes are a result of spot evolution, sometimes it looks
like a new spot or spot group is appearing or disappearing over the
campaign. It is likely that if we had a longer campaign, then these
objects would be instead in one of the complex peak/resolved peak
categories, but 72 days is not long enough to distinguish the
frequencies.}
\label{fig:shch}
\end{figure}

Examples are seen in Figure~\ref{fig:shch} (also see
Fig.~\ref{fig:mw-moving} for the moving double-dips, which are also
shape changers). Sometimes the overall envelope of the LC changes,
manifesting as changes in shape of the phased LC. In other cases, one
can see additional structure appear in the LC and grow in strength, or
even move with respect to the other structures in the LC over the
course of the campaign. 

Spot evolution seems like a likely physical interpretation; one can
imagine a new spot or spot group appearing or disappearing over the
campaign to create these LCs.  For those where structure in the LC
moves with respect to other structures in the LC, spot evolution
combined with latitudinal differential rotation could explain the
observations. We note, though, that there are other LC categories
where we find two or more periods that we attribute to latitudinal
differential rotation and/or spot evolution. In the case of shape
changers, the changes are evidently happening more slowly, such that
there are not enough rotation periods (and/or stability in the
periods), and so an additional period cannot be resolved.  If we were
able to obtain light curves over more than 72 days, perhaps a second
(or third) period could be derived, in which case some of these stars
could end up in the `complex peaks' and/or `beater' categories above.

The shape changers have 1$\lesssim$\vmk$\lesssim$5. While the \vmk\ range
for shape changers overlaps that for beaters and complex peaks, the
shape changers extend to redder \vmk; see additional discussion below
and in Paper III. About 60\% of the shape changers are also in another
category described above. 

\subsection{Orbiting Clouds?}
\label{sec:debris}
\label{sec:batwings}

\begin{figure}[ht]
\epsscale{0.9}
\plotone{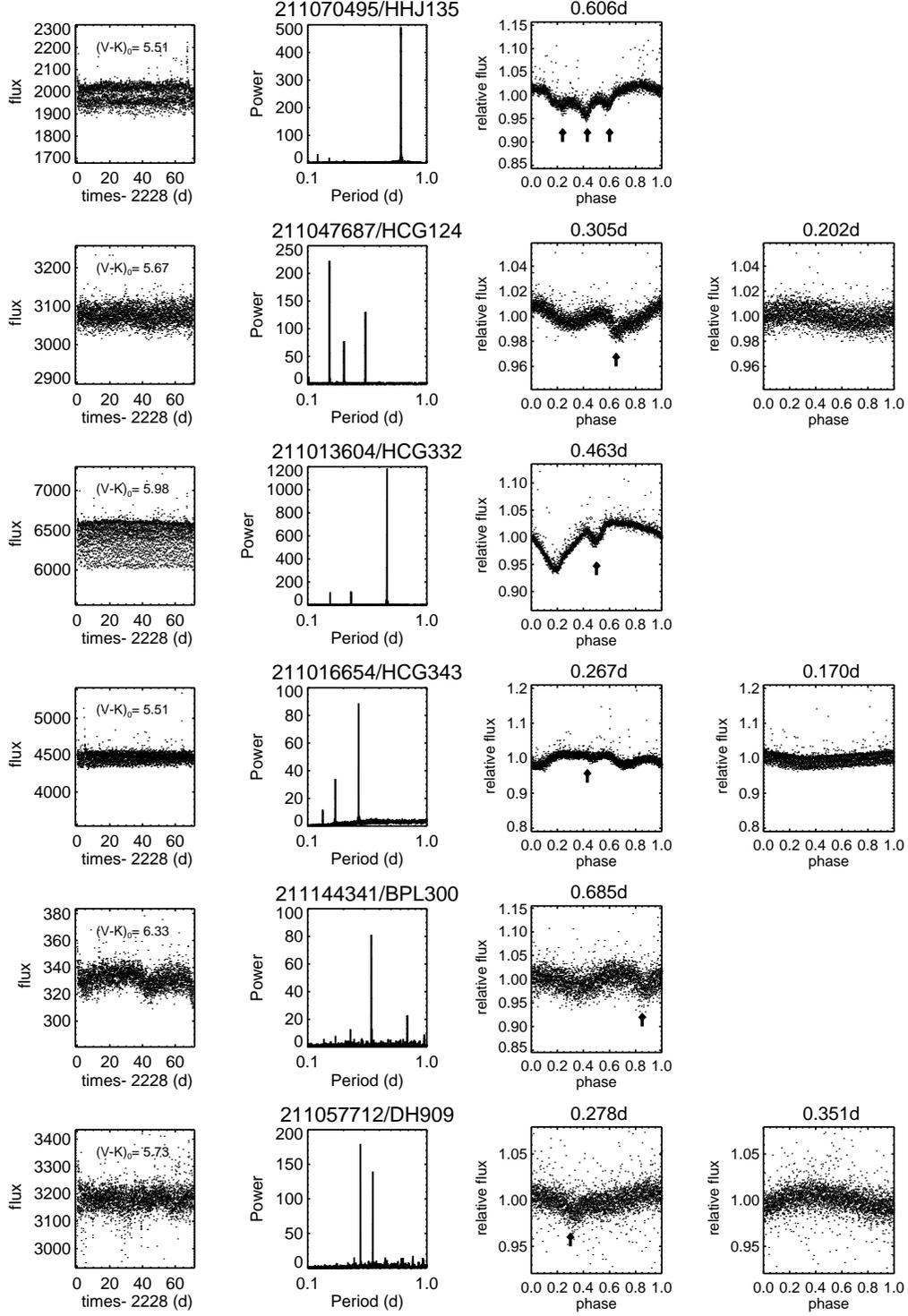}
\caption{All six stars identified as having repeated, angular, brief
dips in their light curves.  First column: full LC (with \vmk\ for star as
indicated); second column:
periodogram; third column: phased LC for maximum peak; fourth column:
phased LC for secondary peak, if present.  Rows, in order: 
211070495/HHJ135, 211047687/HCG124, 211013604/HCG332,
211016654/HCG343, 211144341/BPL300, 211057712/DH909. Arrows in the
first phased LC indicate the shallow, angular dips discussed in
Sec.~\ref{sec:batwings}. These LC structures may be due to orbiting
clouds or debris.}
\label{fig:batwings}
\end{figure}

In six cases, the period we identify in the power spectrum yields a
phased light curve of unusual shape; see Fig.~\ref{fig:batwings}.  The
phased light curves show shallow, angular dips (highlighted with
arrows in Fig.~\ref{fig:batwings}) which cover a relatively small
portion of the total phase.   The pattern is stable (or nearly stable)
over the 72 day K2 campaign. In several cases, the light curves also
show broad dips, which presumably are due to cool spots on the stellar
surface. Half of these stars have an additional secondary period. 
Four are high-quality members, one is a lower quality member
(211013604/HCG332) and one is a probable non-member
(211144341/BPL300).   About 0.6\% of the sample (0.7\% of the periodic
sample) have these structures.

The narrowness of the highlighted dips seem impossible (or at least
very difficult) to explain with cool spots. A Fourier decomposition of
a star spot light curve has essentially no power beyond the second
harmonic. Foreshortening and limb darkening conspire to ensure that
for spots, there are never more than two humps per light curve. The
variations in these stars are too rapid to be fitted with a linear
combination of $P$, $P/2$, and $P/3$. However, these dips are also
too broad to be due to transits of secondary stars or giant planets.  
The basic information for these stars is all in
Table~\ref{tab:bigperiods}; all of them are fairly late M dwarfs, and
all have periods $<$ 0.7d, which means they have fairly normal periods
for Pleiades stars of this mass.  

We have identified two possible physical explanations for the shallow
dips in these light curves.  First, they could be due to eclipses of
warm, dense circumstellar clouds formed near the top of large coronal
loops and temporarily held near the Keplerian co-rotation radius by
centrifugal force (Collier Cameron 1988; Collier Cameron \& Robinson
1989).   Such clouds have been identified in synoptic spectra of
several rapidly rotating, young dwarfs (\eg, AB Dor), though not
previously in time series photometry.   For at least four of the six
stars, the cloud associated with the narrow dip would be located near
the corotation radius since the periods for the narrow and broad dips
are apparently the same, or nearly the same. In two cases, the narrow
dip appears to move slightly in phase relative to the broad dip or to
another narrow dip; differential rotation is very unlikely in these
stars, given their spectral type and rotation rate. To see slingshot
prominences in broadband photometry, we would have to be observing
eclipses of continuous bound-free emission as the cloud passes behind
the star. For this reason, we would only expect to see the phenomenon 
easily in rapidly rotating stars with very low photospheric surface
brightnesses, \ie, late M dwarfs. 

The other possible physical mechanism for the narrow dips is that they
could be due to comets or some other form of ``debris" orbiting these
stars, perhaps similar in some respects to stars from the main Kepler
field recently found to show transient, shallow flux dips (Boyajian
\etal\ 2016; Vanderburg \etal\ 2015).  In this case, the dips would be
associated with transits of these dusty structures in front of the
star.   It is not obvious why material would be located near the
Keplerian co-rotation radius so frequently in this model.

If these objects have orbiting particulate debris, one might expect an
IR excess from the debris. However, none of these objects have truly
compelling evidence for an infrared (IR) excess; the longest
wavelength detection for any of them is WISE-3, 12 \mum, which is
detected in four of the six stars. The most likely excess is found in
211013604/HCG332, which has [3.4]$-$[12]=0.53 mag and the metric often
used to assess signficance of excess,
$\chi=([3.4]-[12])/\sqrt{\sigma^2_{[3.4]}+\sigma^2_{[12]}}$, is 5, so
not very obvious. The other stars with 12 \mum\ detections have
similar sized [3.4]$-$[12] but less significant $\chi$.  Moreover,
only a small fraction of dust could account for these dips; see, \eg,
Gillen \etal\ (2014), Terquem \etal\ (2015), and Gillen \etal\
(2016).  Further investigation is warranted, such as monitoring in
different wavelengths, and a more careful assessment of any small IR
excesses.

HCG332 is one of two stars originally made famous in Oppenheimer
\etal\ (1997). This work reported results of a search for lithium in
very low mass members of the Pleiades; they were attempting to
identify stars at the lithium depletion boundary marking the point
cooler than which Pleiades age stars have not reached core
temperatures hot enough to burn lithium.  Oppenheimer \etal\ found no
stars fitting the criteria to be at the lithium depletion boundary,
but they did identify two purported, somewhat higher mass Pleiades M
dwarfs that had strong lithium features when no lithium should be
present. (The other star is HCG509, discussed in Paper I's Appendix.)
Both stars are located $>$1 mag above the Pleiades single-star main
sequence, which they interpreted to mean they are very young
pre-main-sequence stars and not Pleiades members. We will report
elsewhere on the implications for models of these two stars based on
their K2 lightcurves and additional spectroscopy (Barrado \etal\ in
prep).


\subsection{Pulsators and Rotation}
\label{sec:pulsators}

Throughout most of the mass range of interest to this paper,
rotational modulation due to star spots provides the only plausible
physical mechanism to explain the periodic LCs we see. 
However, at the high mass end of our sample (\vmk$\lesssim$1.3,
$M\gtrsim$ 1.2 \msun), pulsation may provide an alternative
explanation. 

\begin{figure}[ht]
\epsscale{1.0}
\plotone{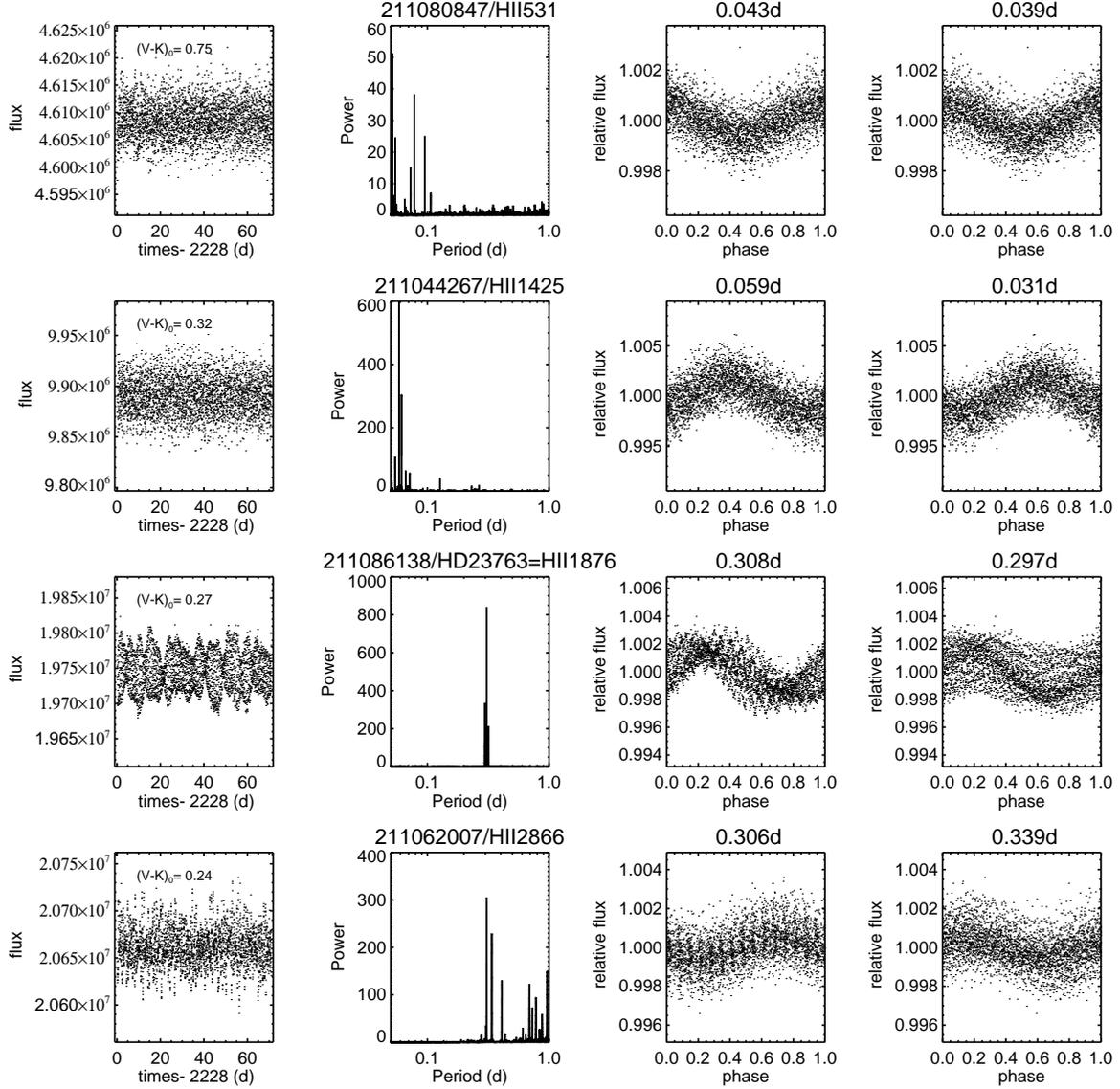}
\caption{Four examples of the pulsator category, stars that are likely
$\delta$ Scuti pulsators.  First column: full LC (with \vmk\ for star as
indicated); second column: LS
periodogram; 3rd column: phased LC at the $P$ with the highest peak in
the periodogram; 4th column: phased LC at the $P$ with the second
highest peak in the periodogram. Rows, in order:  EPIC
211080847/HII531,  211044267/HII1425, 211086138/HD23763=HII1876,
211062007/HII2866.   }
\label{fig:pulsators}
\end{figure}

In eight cases ($\sim$1\% of the sample), the power spectrum reveals a
`forest' of significant (FAP=0) peaks in the periodogram, all at very
short periods; see Figure~\ref{fig:pulsators} for examples and
Table~\ref{tab:pulsators} for a list of these stars. They are also all
very small amplitude LCs (median 0.0027 mag). In most cases, these
must be pulsators; if interpreted as rotation periods, the period
would exceed the break-up limit.  They are all earlier types -- 
their \vmk\ ranges from 0.24 to 0.85 (mid-A to early
F). Five of the eight stars we have identified in this category have
periods $\lesssim$0.1d; the other three are all $\sim$0.3 d and
include both the bluest and reddest \vmk\ of this category (spectral types A1,
A2, and A9; see Table~\ref{tab:pulsators}).

Delta Scuti stars have spectral types A or F, and periods typically
$\lesssim$0.3 d (\eg, Breger 1979). They are lower mass Cepheid
analogues.  Several of the Pleiades with K2 LCs have been identified
in the literature (Breger 1972, Fox Machado \etal\ 2006) as being
$\delta$ Scuti pulsators, but most of them are too bright for us to
reliably identify a period at all with the light curve versions we
have, much less a pulsator-type power spectrum. Two of them are ones
we identified as pulsators, and the rest are very bright; see
Table~\ref{tab:pulsators}. For two of these bright stars,  knowing
that they are thought to be pulsators, one can find the likely
relevant peaks in the power spectrum in amongst the noise, but the
light curve is significantly compromised and we did not identify these
periods {\em a priori}.  


\begin{deluxetable}{cccp{9cm}}
\tabletypesize{\scriptsize}
\tablecaption{Pulsators\label{tab:pulsators}}
\tablewidth{0pt}
\tablehead{\colhead{EPIC} & \colhead{Other name}& \colhead{SpTy}
&\colhead{Notes} }
\startdata
211018096&HD23791=HII1993 & A8&  Identified {\em a priori} by us as a pulsator, $P\lesssim$0.1d; obvious $\delta$ Scuti \\
211044267&HII1425& A3V &Identified {\em a priori} by us as a pulsator, $P\lesssim$0.1d; obvious $\delta$ Scuti; literature-identified $\delta$ Scuti (Breger 1972, Fox Machado 2006); in Fig.~\ref{fig:pulsators} \\
211057064&HD23863=HII2195&A7V&Identified {\em a priori} by us as a pulsator, $P\lesssim$0.1d; obvious $\delta$ Scuti \\
211066615&HII652&A3V& Identified {\em a priori} by us as a pulsator, $P\lesssim$0.1d; obvious $\delta$ Scuti \\
211080847&HII531 &A9m\tablenotemark{a}& Identified {\em a priori} by us as a pulsator, $P\lesssim$0.1d; obvious $\delta$ Scuti; in Fig.~\ref{fig:pulsators} \\
211062007&HII2866 &A2V& Identified {\em a priori} by us as a pulsator, $P\sim$0.3d; likely $\delta$ Scuti; in Fig.~\ref{fig:pulsators}\\
211086138&HD23763=HII1876 &A1V&Identified {\em a priori} by us as a pulsator, $P\sim$0.3d; likely $\delta$ Scuti; in Fig.~\ref{fig:pulsators}\\
211093705&HII697& A9& Identified {\em a priori} by us as a pulsator, $P\sim$0.3d; likely $\delta$ Scuti; literature-identified $\delta$ Scuti (Breger 1972)\\ 
\hline
211115721&HII1266 &A9V&Literature-identified $\delta$ Scuti (Fox Machado \etal\ 2006), not identified by us, for which we have just one (long) period\\
211072836&HII1362 &A7&Literature-identified $\delta$ Scuti (Fox Machado \etal\ 2006), not identified by us because star too bright\\
211088007&HII158 &A7V&Literature-identified $\delta$ Scuti (Fox Machado \etal\ 2006), not identified by us because star too bright (though knowing that it is a pulsator, one can identify the likely relevant periodogram peaks)\\
211101694&HII1384&A4V&Literature-identified $\delta$ Scuti (Fox Machado \etal\ 2006), not identified by us because star too bright (though knowing that it is a pulsator, one can identify the likely relevant periodogram peaks)\\
\enddata
\tablenotetext{a}{All of the spectral types here come from Mendoza
(1956), except for this one, HII531, which Mendoza (1956) lists as
``Am?," but Gray \etal\ (2001) give a more definitive A9m.}
\end{deluxetable}

We suspect that the remaining stars we identified as likely pulsators
are also $\delta$ Scutis, particularly the three with the shortest
periods. The other three near $\sim$0.3 d are at the outer edge of the
accepted range for $\delta$ Scuti periods, as well as the outer edge
of the expected spectral type range. However, given the spectral types
for the stars already in the literature as $\delta$ Scutis (see
Table~\ref{tab:pulsators}), perhaps the spectral type range for these
pulsators is not so rigidly defined. We thus identify all six of our
{\em a priori} identified pulsators as $\delta$ Scutis. No modern
\vsini\ values are available for these stars in the literature.

The $\delta$ Scutis are relatively easily identified due to their very
short periods that essentially cannot be ascribed to rotation because
the inferred rotational velocities would exceed breakup.  However,
another type of pulsating variable is the $\gamma$ Dor type 
(\eg, Krisciunas 1994; Kaye \etal\ 1999), which may have the same
underlying physics as $\delta$ Scutis (\eg, Xiong \etal\ 2016).  Gamma
Dors are also A or F stars, but are pulsating with on average longer
periods, $\sim$0.4 to 3 days (\eg, Balona \etal\ 1994, Kaye \etal\
1999). These are harder to distinguish from rotation periods, because
their periods overlap with expected spot-modulated rotation periods
for their masses. 

It has been shown that for stars identified as $\gamma$ Dors in the
main Kepler field, the measured periods correlate well with \vsini\
values for those stars (Balona \etal\ 2011) and similarly for $\gamma$
Dors not necessarily in the Kepler field (Kahraman Ali\c{c}avu\c{s}
\etal\ 2016).   This suggests that it might be difficult to
discriminate between pulsation and rotation as the drivers of the
observed variability for at least some of our F stars.   In fact,
Balona \etal\ (2011) acknowedge that many of the stars they identify
as $\gamma$ Dor variables in the main Kepler field may instead owe
their variability to rotation and star spots. Bradley \etal\ (2015),
despite selecting targets specifically to look for $\gamma$ Dor and
$\delta$ Sct stars in the instability strip in the Kepler field, find
that 74\% of their sample are better interpreted as spot modulation.

\begin{figure}[ht]
\epsscale{1.0}
\plotone{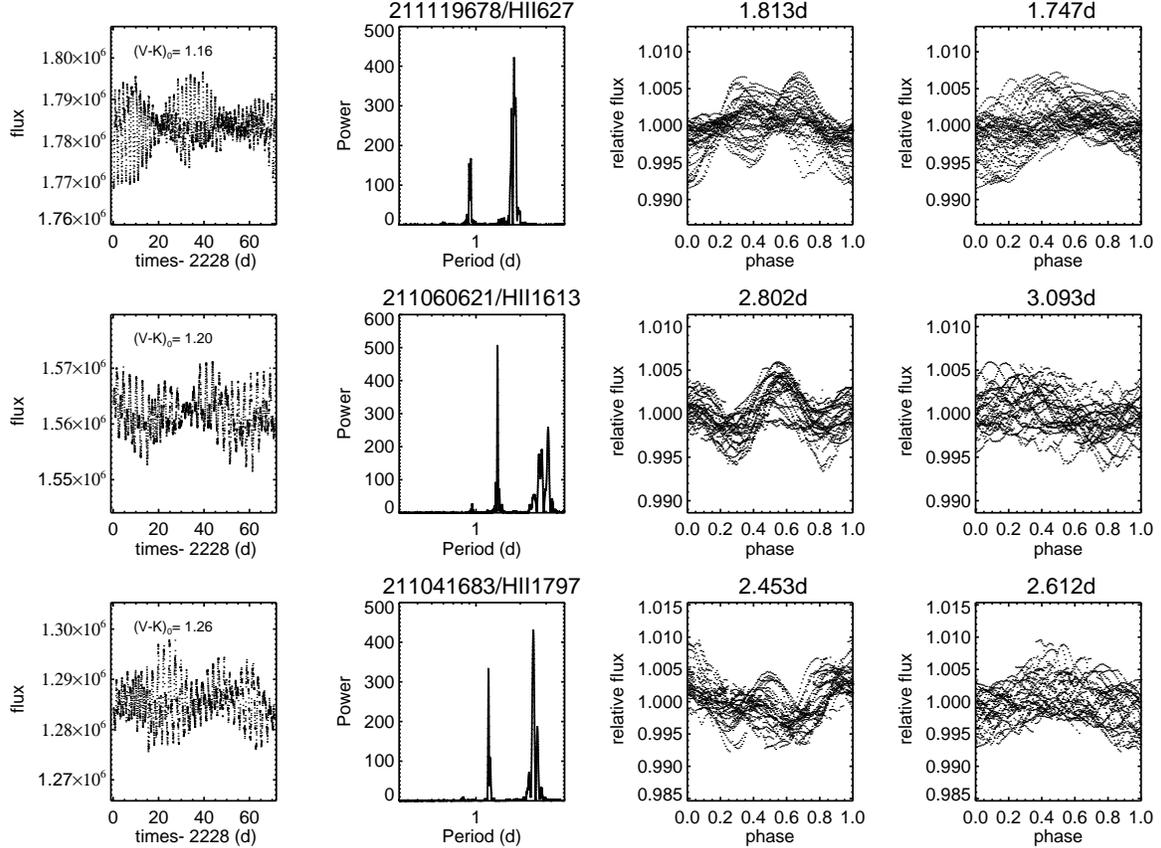}
\caption{Examples of objects that may be  $\gamma$ Dors. 
First column: full LC
(with \vmk\ for star as indicated); 2nd column: periodogram; 3rd
column: phased LC for primary peak; 4th column: phased LC for
secondary peak. Rows, in order: 211119678/HII627, 211060621/HII1613,
211041683/HII1797.  The measured \vsini\ from the literature is
$\sim$33, 20, 20 km s$^{-1}$, respectively, which is consistent with
these being rotation rates.  210990525/PELS124 from
Fig.~\ref{fig:complexpeak} may also be in this category. (Its
\vsini$\sim$20 km s$^{-1}$ as well.) }
\label{fig:gammadors}
\end{figure}

Our close resolved and/or complex peak categories contain many power
spectra that resemble those of $\gamma$ Dors from the main Kepler field in
Balona \etal\ (2011), but ours extend over a much wider range of color than
for just near the instability strip, down to \vmk$\sim$4 for the complex
peaks alone.   There is no clear separation of the LC characteristics of
these stars from slightly more massive or less massive stars, instead there
is simply a smooth transition.  There exist stars with this LC/periodogram
morphology both redward and blueward of the boundaries that are expected to
define the $\gamma$ Dor class.  Figure~\ref{fig:gammadors} includes the LCs
and periodograms for some of the stars with $\gamma$ Dor-like variability.
There is a measured \vsini\ for each of those stars; for the measured $P$,
the measured \vsini\ should be $\sim$20-30 km s$^{-1}$, which they are.  
For this paper, we choose to adopt the dominant LS period as the star's
rotation period for all of these stars.  Given the good correlation between
\vsini\ and period found by Balona \etal\ (2011) and Kahraman
Ali\c{c}avu\c{s} \etal\ (2016), this period is unlikely to be very far from
the true rotation period even if pulsation is the actual cause.  Based on
the wide spectral type range over which we see this type of variability and
the smooth change in LC morphology as one traverses the G to F star
spectral range, we believe it is likely that rotation is the correct
physical mechanism to explain these kinds of LCs and periodograms,
particularly for the ones with \vmk$>$1.1.


\section{Single-Period and Multi-Period Distributions}
\label{sec:singlemulti}

The broadest two categories of LC properties are simply single-period
and multi-period stars. This section looks at where the single-period
and multi-period stars fall in a variety of parameter spaces.

\subsection{Period Distribution}

\begin{figure}[ht]
\epsscale{0.6}
\plotone{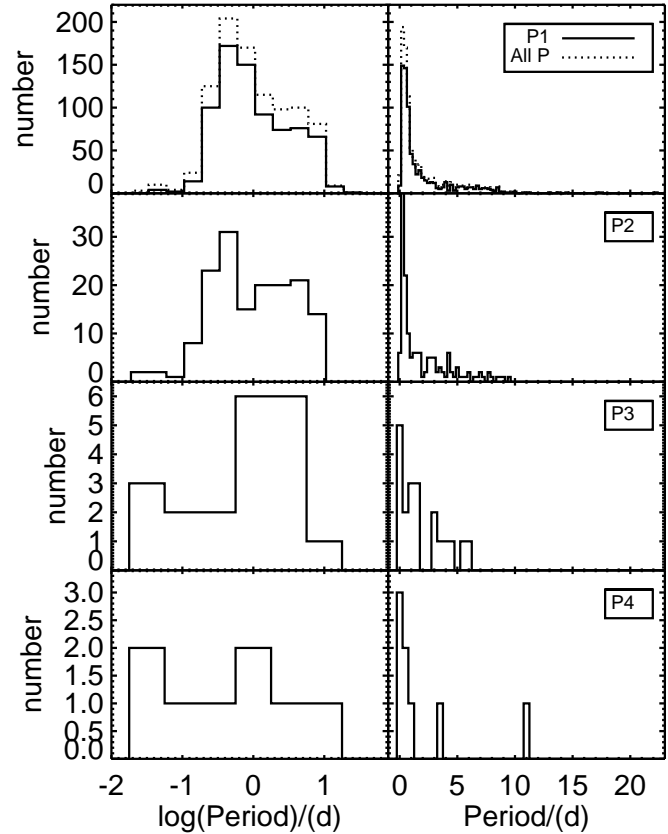}
\caption{Histograms of the periods found by our analysis, in days.  The
left column contains histograms of the log ($P$) and the right column
contains histograms of linear $P$. In the first row,  the solid line
is the primary period (that which we take to be the rotation period of
the star), and dotted line is (for reference) a histogram of {\em all}
the periods found here, including the secondary, tertiary, and
quaternary periods.  The second, third, and fourth rows are the
secondary, tertiary, and quarternary periods plotted separately; there
are many fewer tertiary and quarternary periods.  The period
distribution of all four periods is still strongly peaked at $<$1d.}
\label{fig:phisto}
\end{figure}

Paper I discussed the overall period distribution. Because the sample
of stars with one $P$ dominates, the addition of all the secondary,
tertiary, and quarternary periods does not change the distribution
very much. Figure~\ref{fig:phisto} breaks out the primary, secondary,
tertiary, and quarternary periods found here separately. Even though
there are far fewer tertiary and quarternary periods (10 with tertiary
periods and 5 with quaternary periods), the distributions of all four
periods are still strongly peaked at $<$1d.

If all of the additional periods we determined are periods of
secondary stars in the unresolved binary, then the `true' distribution
of Pleiades rotation rates is the amalgamation of all the periods we
determined, seen as the dotted line in Fig.~\ref{fig:phisto}. However,
it is unlikely that {\em all} of the additional periods are binaries;
it is very likely that some are pulsation and many are differential
rotation and/or spot evolution.

\subsection{\vmk\ Colors}

\begin{figure}[ht]
\epsscale{0.7}
\plotone{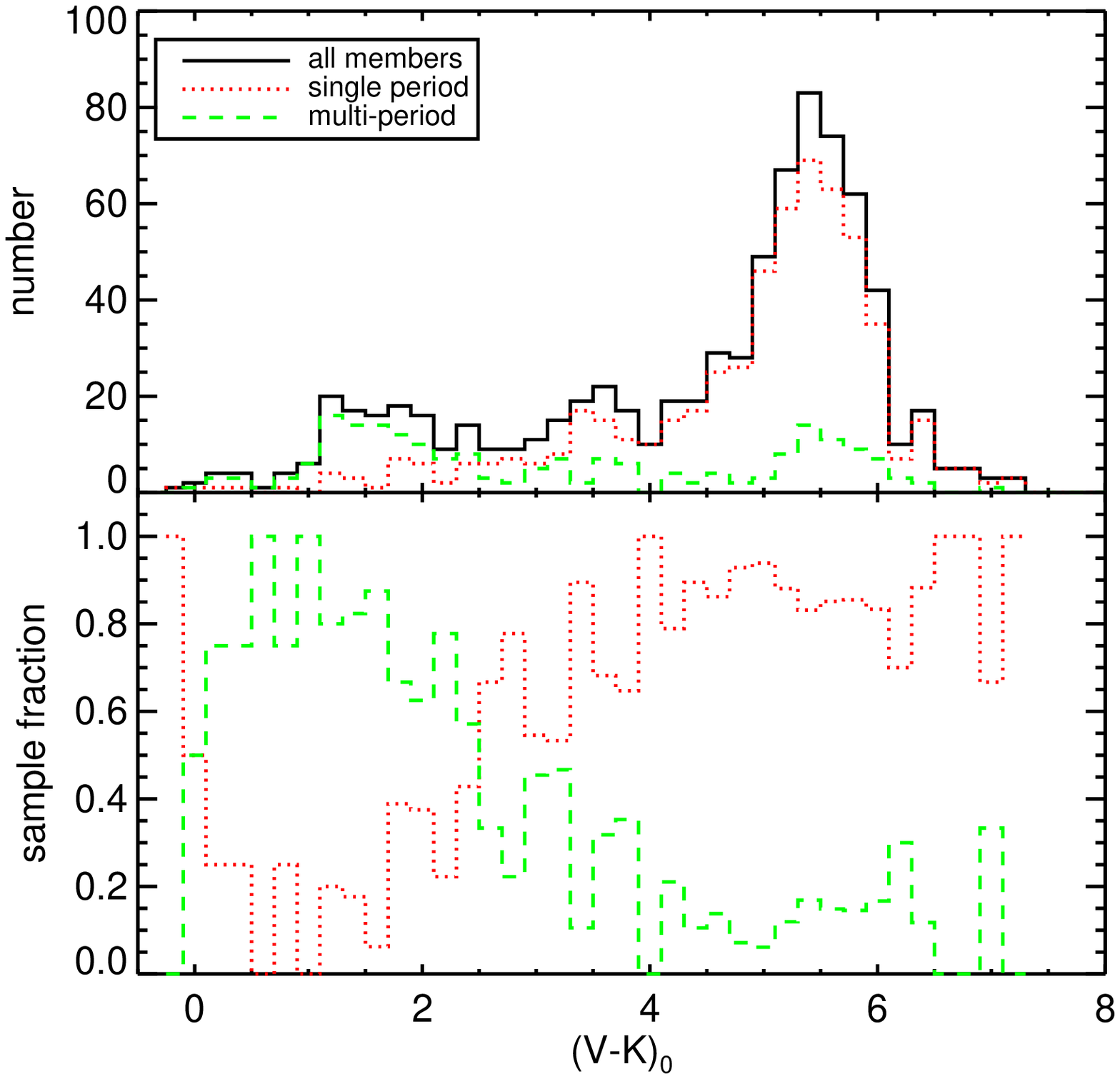}
\caption{Distribution of \vmk\ for the ensemble, with the single (red
dotted) and multiple (green dashed) populations called out. The top
panel is absolute numbers, and the botttom is the sample fraction. The
earlier types (bluer colors) are preferentially multi-periodic, and
the later types (redder colors) are preferentially single-period,
though about 30\% of the earlier types have only one period that we
can detect, and about 20\% of the later types have multiple periods.
Note that the place where the single and multi-period distributions
swap dominance in the lower panel is at about \vmk$\sim$2.6, the
location of the `kink' discussed in Paper III. }
\label{fig:vkdistrib}
\end{figure}

Figure~\ref{fig:vkdistrib} shows the distribution of \vmk\ values for
the single-period and multi-period categories. The earlier types
(bluer colors, G \& K dwarfs) are preferentially multi-periodic, and
the later types (redder colors, M dwarfs) are preferentially
single-period, though about 30\% of the earlier types have only one
period that we can detect, and about 20\% of the later types have
multiple periods. Many of the earlier types categorized as having only
one period also have broad periodogram peaks, suggesting that they
really are multi-periodic; we just cannot ascertain exactly what
that additional frequency is. 

The relationship found in Fig.~\ref{fig:vkdistrib} could be explained
by the ratio of active region lifetime to rotation period. If the
ratio is large, the LC is coherent even though it may have
substructure. If the ratio is small, spots live for only a small
number of rotations, giving rise to phase and amplitude modulation and
peak splitting in the periodogram. If spots decay through diffusion,
and the diffusivity is related to the convective velocity, cooler
stars will have longer lived spots. 

\subsection{Amplitude Distribution}

\begin{figure}[ht]
\epsscale{0.6}
\plotone{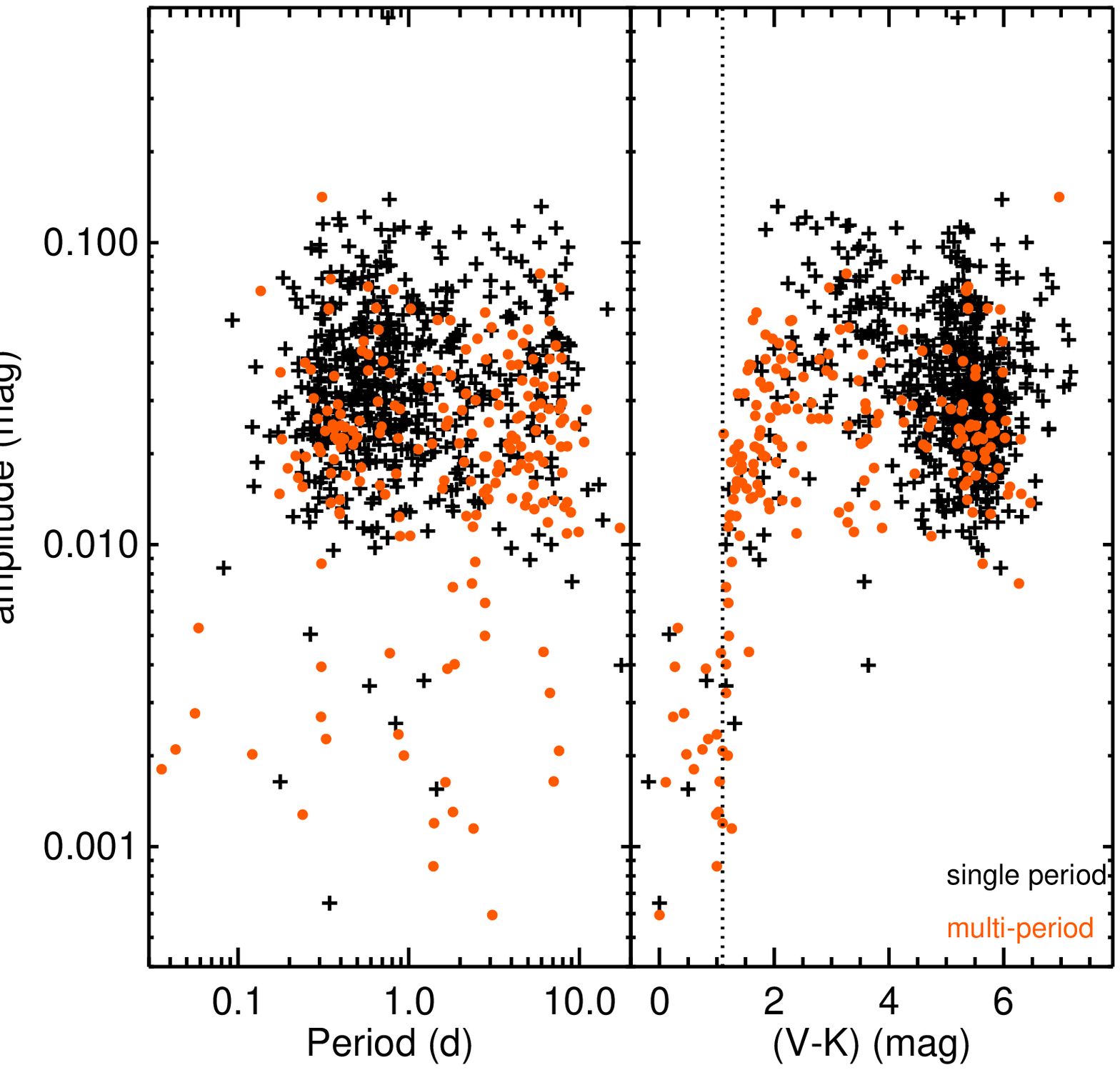}
\caption{The amplitude (from the 10th to the 90th percentile), in
magnitudes, of the periodic light curves, against $P$ and \vmk. The
stars with single periods are black crosses and those with multiple
periods are orange dots. The vertical dotted line is at \vmk=1.1 (see
Paper III, where there is a linear version of this plot). 
Stars bluer than about \vmk$\sim$1.1 have clearly lower amplitudes.  }
\label{fig:ampldistrib}
\end{figure}

The LCs with multiple periods are on average slightly lower amplitude
than those with single periods; see Figure~\ref{fig:ampldistrib} (and
see Paper I, section 3.3 for the definition of amplitude employed
here). For the single period sample, the mean amplitude (10-90\% of
the points fall within this range, in magnitudes) is 0.0399$\pm$0.0321
mag. For the multi-period sample, the mean is 0.0250$\pm$0.0183 mag.
Considering just \vmk$>$1.1 to remove the possible and likely
pulsators (also see discussion in Paper III), the single period sample
is nearly unchanged at 0.0403$\pm$0.0321 mag and the multi-period
sample is 0.0274$\pm$0.0176 mag; the multi-period amplitudes are still
on average smaller than the single-period amplitudes. Removing the
distant resolved peaks as likely binaries (where the amplitude may be
diluted by the companion) makes little difference in the means because
there are relatively few of them.  There are essentially no
multiperiodic stars with amplitudes above 0.08 mag, but the stars with
single frequencies include many up to 0.13 mag.  This is likely why
little if any of this diversity of multi-period stars in the Pleiades
has been discovered prior to these K2 data. (We note, however, two
items of relevance: Stars with beating light curves have been observed
in non-Pleiades CoRoT and Kepler data, such as Nagel, Czesla, \&
Schmitt (2016). Magnitskii (2014) identified more than one period for
211025716/HII296, interpreting it as substantial motions of spots in
latitude and longitude; this star's K2 LC shows evidence of beating
and has a complex periodogram peak, though we can only derive one
period for it.) 

The bias towards our finding single periods at later types is probably
partially a selection effect (lower signal-to-noise available in the
fainter later types), but may also be due to different dynamo regimes
or different spot latitude distributions. If the multi-periodic stars
originate from spots at different latitudes (\eg, differential
rotation), then the later types, which are largely rapidly rotating,
are either rotating as solid bodies or primarily have high latitude
spots. We discuss this more below and in Paper III.

\subsection{$P$ vs.\ \vmk}

\begin{figure}[ht]
\epsscale{0.8}
\plotone{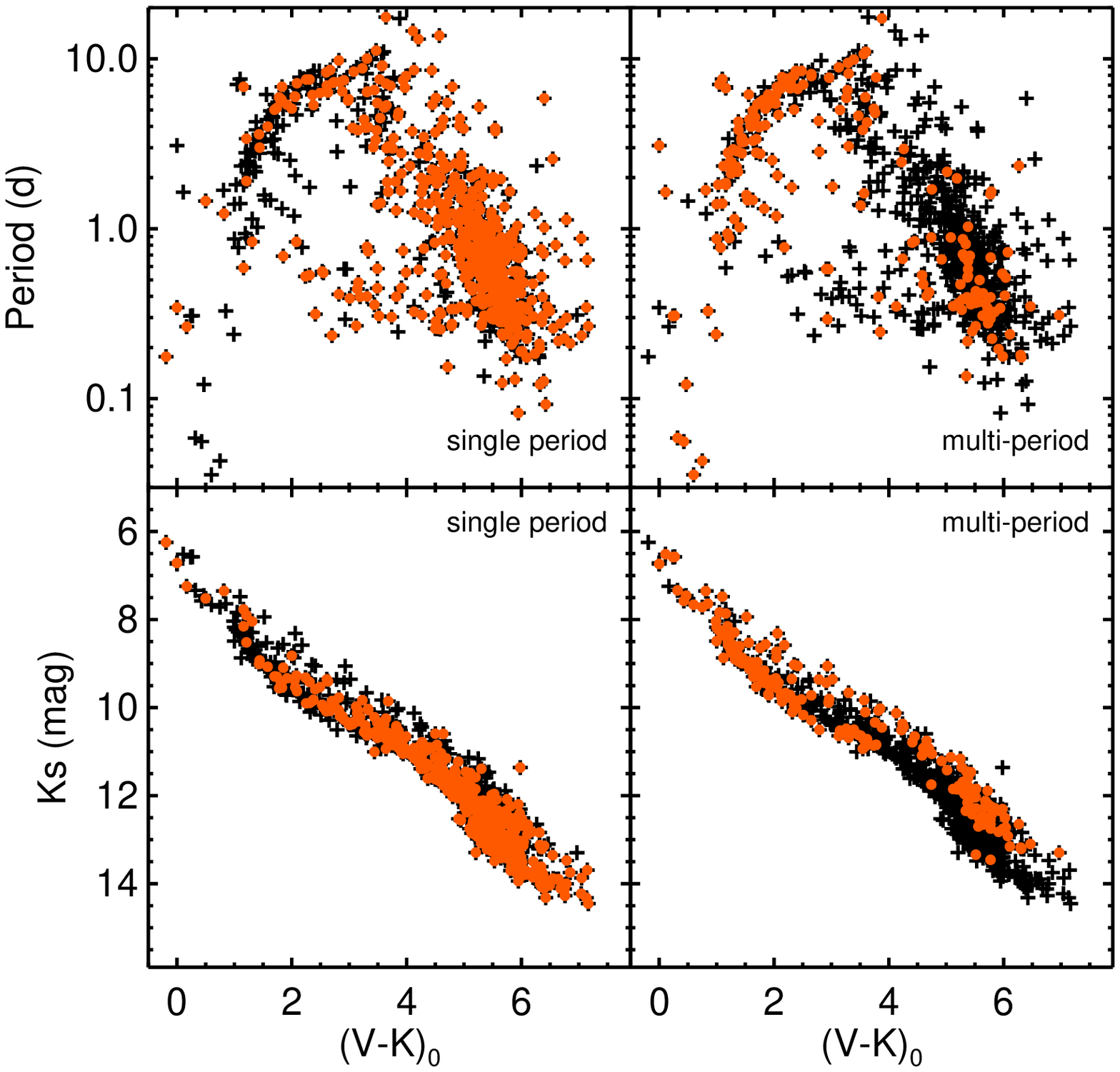}
\caption{Plot of $P$ vs.~\vmk\ (top row) and \ks\ vs.~\vmk\ (bottom
row) highlighting the single-period (left) and multi-period (right)
populations. Most of the multi-period stars are earlier-type stars,
and the slow sequence is dominated by multi-period stars. The fast
sequence is dominated by single-period stars. Those stars in the fast
sequence that are multi-periodic also tend to be the photometric
binaries, suggesting that most of the M stars have single
periods, and may therefore be rotating as solid bodies.   }
\label{fig:pvmk3freq}
\end{figure}

Fig.~\ref{fig:pvmk3freq} shows the locations of the single and
multi-period stars in the $P$ vs.~\vmk\ diagram and the
color-magnitude diagram (CMD) for reference.  As we suspected from
Fig.~\ref{fig:vkdistrib}, it is primarily the earlier types that have
multiple periods, and the later types that have single periods. Most
of the stars in the slowly rotating sequence with
$1.1\lesssim$\vmk$\lesssim 3.7$ have multiple periods. Most of the rest
of the  $P$ vs.~\vmk\ diagram, particularly the M stars in the fast
sequence (with \vmk$\gtrsim$5.0), is composed of single-period stars.
Interestingly, the M stars that are multi-period in the fast sequence
tend to also be above the main sequence, \eg, photometric binaries.
For stars redder than \vmk$\sim$4, nearly all the multi-period LCs
are also photometric binaries. Presumably, then, in those cases, the
two periods we derive from the LC come from two different stars. Since
most of the rest of the stars in the fast sequence seem to have single
periods, and the two components of the photometric binary each have
single periods, we suggest that most of the M stars have single
periods, and may therefore be rotating as solid bodies, or the
timescale of any shift in latitude is much longer than the K2
campaign.


\section{Distributions of Categories}
\label{sec:alltherest}

While the prior section investigated where single- and multi-period
stars fall, we have many more than just two categories of LCs and
periodogram structures; this section uses the full range of categories
to explore the possible physical interpretations of the LC and
periodogram shapes.

\begin{figure}[ht]
\epsscale{0.8}
\plotone{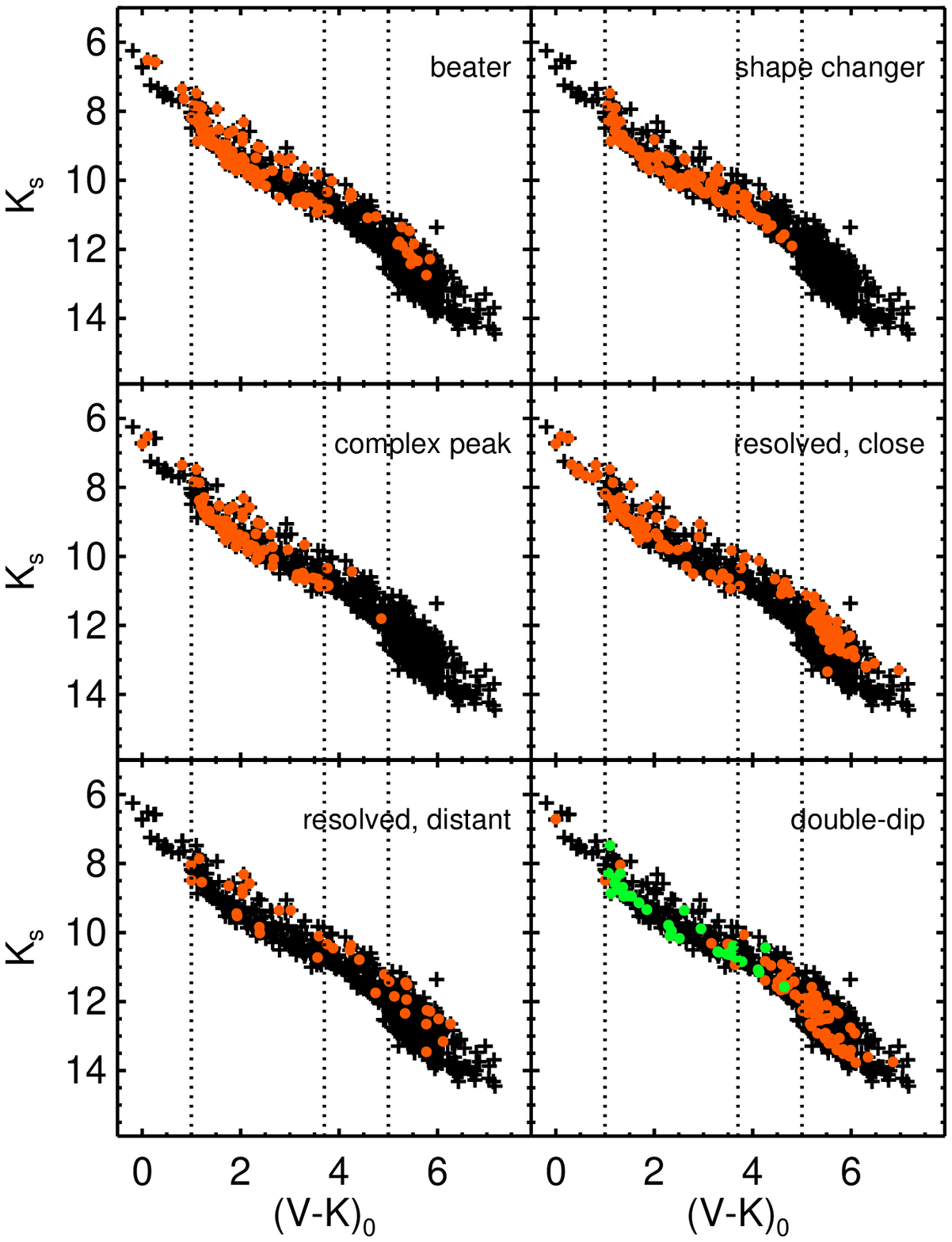}
\caption{Plot of \ks\ vs.~\vmk, highlighting several of the LC
categories described in the text. In each case, black crosses are the
ensemble, and colored points highlight those objects of the category
indicated. In the last panel, orange dots = stationary double-dip, and
green dots = moving double-dip. The bluer stars (slow sequence)
include most of the beaters, shape changers, and complex peaks, and a
significant number of the resolved close peaks. Resolved close peaks
are also found in the redder stars (fast sequence), and these are
largely the photometric binaries from Fig.~\ref{fig:pvmk3freq}. Some
beaters and resolved distant peaks are also photometric binaries.  The
double-dip LCs above are found throughout this diagram, but those showing
movement within the campaign are distinctly higher mass. The dotted
vertical lines are at \vmk=1.1, 3.7, and 5.0, the divisions between the
F stars ($<$1.1), slow sequence (1.1 to 3.7), the `disorganized
region' (3.7 to 5.0), and the fast sequence ($>$5.0); see
Papers I and III.}
\label{fig:optcmdclasses}
\end{figure}

\begin{figure}[ht]
\epsscale{0.8}
\plotone{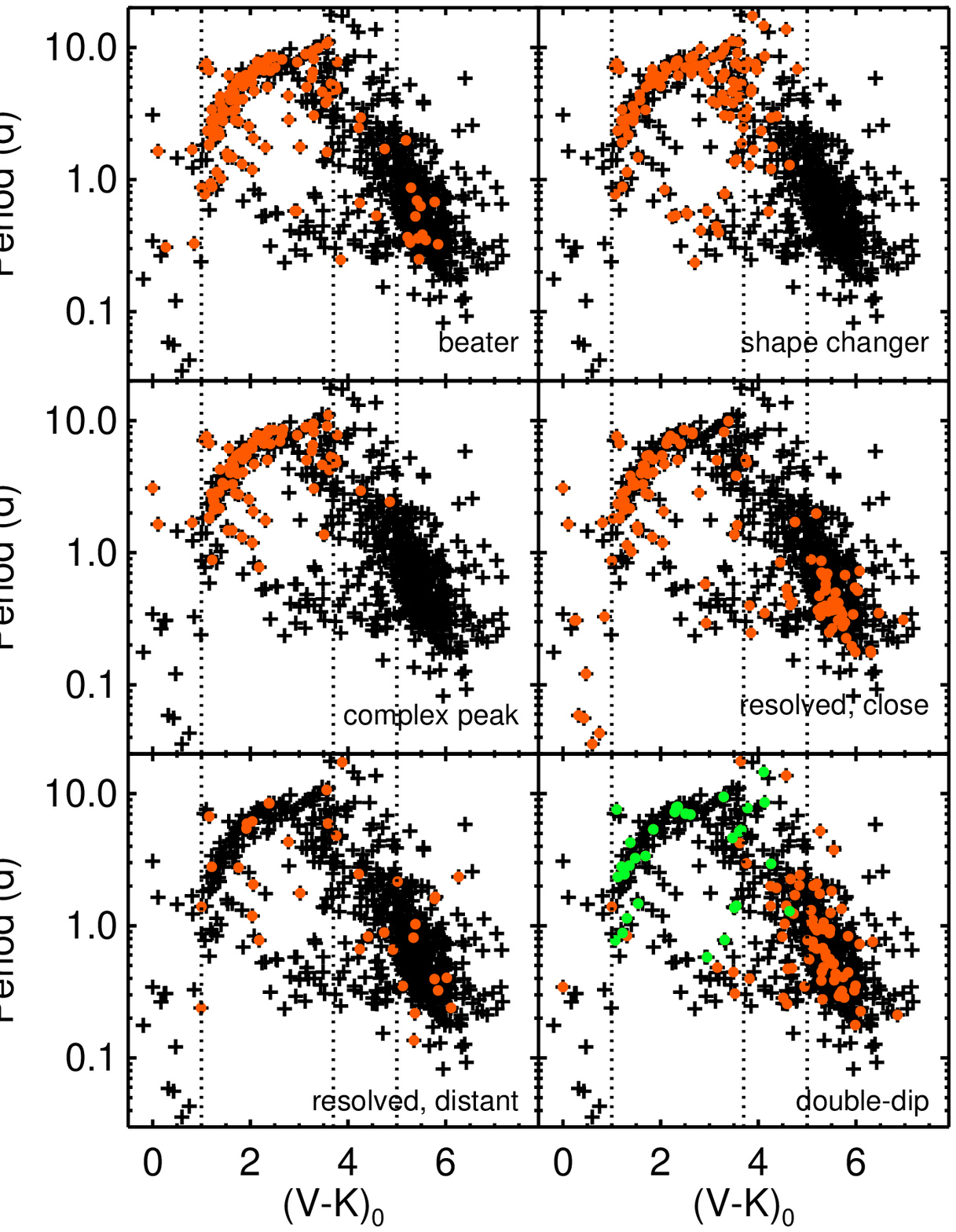}
\caption{Plot of $P$ vs.~\vmk, highlighting several of the LC
categories described in the text. Notation (and the location of the
dotted vertical lines) is same as
Fig.~\ref{fig:optcmdclasses}. 
The slow sequence
consists of beaters, shape changers, complex peaks, and resolved close
peaks. Resolved close peaks are also found in the fast sequence, and
these are largely the photometric binaries from
Fig.~\ref{fig:pvmk3freq}. Some beaters and resolved distant peaks are
also photometric binaries.  The double-dip LCs above are found through
this diagram, but those showing movement within the campaign are
distinctly higher mass.   }
\label{fig:pvmk4classes}
\end{figure}

\subsection{Beaters, Shape Changers, and Complex Peaks}

We suggested above that beaters, shape changers, and complex peaks may
all be different manifestations of more or less the same physics, that
of latitudinal differential rotation and/or spot evolution. The
beaters, shape changers, and complex peaks are highlighted separately
in the CMD in Fig.~\ref{fig:optcmdclasses} and the $P$ vs.\ \vmk\
diagram in Fig.~\ref{fig:pvmk4classes}. All of these classes dominate
the slow sequence at $1.1\lesssim$\vmk$\lesssim 3.7$. The beaters make a
larger excursion down into the fast rotating sequence with
\vmk$\gtrsim 5.0$ than the shape changer or complex peak classes, and
the shape changers seem to dominate the `disorganized' region with
$3.7\lesssim$\vmk$\lesssim 5.0$. This is consistent with various
manifestations of spot/spot group evolution and/or differential
rotation. Especially for those shape changers (like
Fig.~\ref{fig:mw-moving} or  Fig.~\ref{fig:shch}) where one can see
the spot evolution on timescales of the K2 campaign, the spot
evolution is so slow, or the latitudinal differential rotation is so
weak that spots at different latitudes have very similar rotation
rates, the K2 campaign is not long enough to distinguish that second
frequency. If the `disorganized region' is a transition from strong
differential rotation/fast spot evolution to solid body rotation/a
more stable spot or spot group, then this is consistent. Moreover,
there seem to be two populations in the `disorganized region', fast
and slow rotators; there is a gap near 1d. The slower rotators are
more likely to be shape changers than the fast rotators, again
consistent with the faster rotators being solid-body rotators.

\subsection{Double-dip LCs}

On the whole, the double-dip LCs above are found throughout
Fig.~\ref{fig:pvmk4classes} (for \vmk$>$1.1), with no particular
preference for color or rotation rate. This is consistent with our
interpretation of them as spots on well-separated longitudes -- the
effect can occur whenever there are spotted stars, with no preference
for color or rotation rate. However, those that show changing shapes
over the K2 campaign (moving double-dip) are distinctly higher mass
than those that show no shape changes (stationary double-dip). This is
consistent with our proposed interpretation of  spot
evolution/migration over the K2 campaign for the moving double-dip.
The lower-mass stars are more likely to be rotating as solid bodies,
but spot lifetime also increases towards lower \teff. 

\subsection{Resolved Peaks}

For the power spectra where we can resolve the specific periods in
either close or distant pairs, there is probably more than one
physical explantion underlying this observed property.

Fig.~\ref{fig:pvmk4classes} shows that the population with close peaks
in the FGK stars seems also to track the slow sequence again
consistent with differential rotation. However, in the fast sequence
of M stars, there are also many more close peaks than the beater/shape
changer/complex peak category.   As can be seen in
Figs.~\ref{fig:optcmdclasses}, by the M stars, essentially all of the
close resolved peaks are likely photometric binaries. This possibility
is explored in more detail in Paper III.

The distant peaks seem to represent a more dispersed population in
both Figs.~\ref{fig:optcmdclasses} and \ref{fig:pvmk4classes}, less
consistent with the large-scale structure in
Fig.~\ref{fig:pvmk4classes}.  Both of these characteristics would be
consistent with the distant peaks being more likely to be binaries as
opposed to differential rotation or spot evolution. The population of
late-type multi-period photometric binaries from
Fig.~\ref{fig:pvmk3freq} are largely the resolved close periods,
though some beaters and resolved distant periods are also photometric
binaries in this color range.  For the distant peak stars, because
the peaks are so widely spaced in period, forcing the peaks to be 
differential rotation would require an unphysically large shear. The
most extreme case is 210930791/BPL72, an M star, with two periods of
$\sim$19 d and $\sim$1.7 d; a more typical period difference for the
resolved distant category is $\sim$1.4d.


\section{$\Delta P$ distributions}
\label{sec:deltaP}

\begin{figure}[ht]
\epsscale{0.8}
\plotone{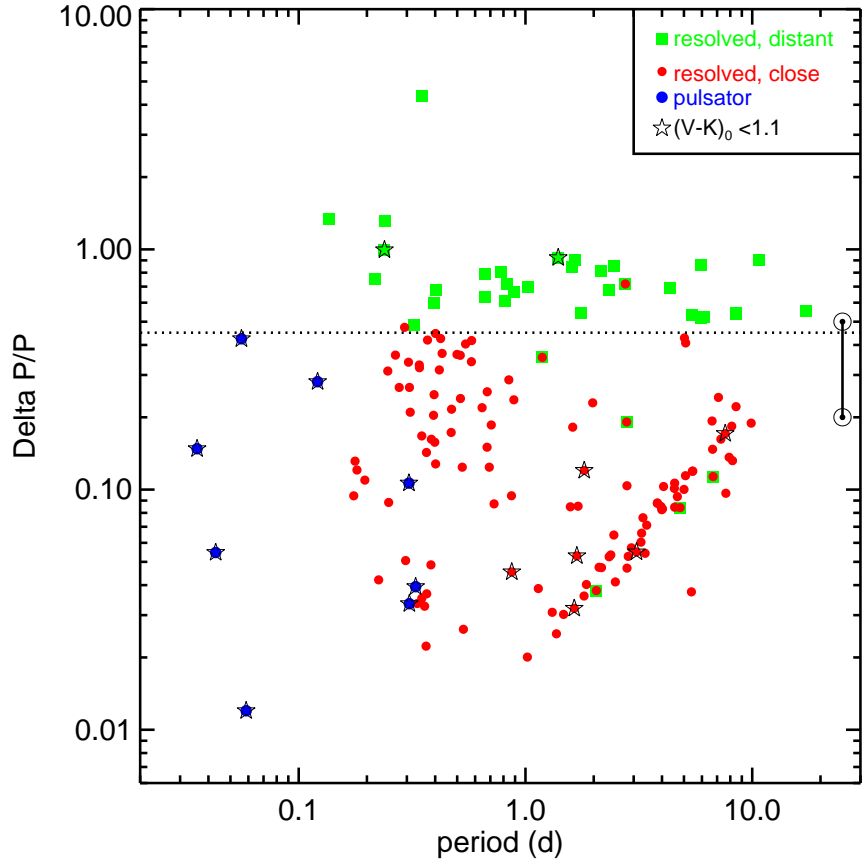}
\caption{Plot of $\Delta P/P_1$ vs.~$P$ for pulsators (blue dots),
resolved distant peaks (green squares), and resolved close peaks (red
dots). An additional black star indicates that those stars have
\vmk$\leq1.1$ (see Paper III). The range of possible values for the 
Sun is included for reference
($\odot$);  if one takes as $\Delta P$ the range of periods measured
where sunspots occur, $\Delta P/P_1 \sim 0.1-0.2$, but if one takes
the full range of $\Delta P$, equator to pole, $\Delta P/P_1 \sim
0.5$. The dotted line is at $\Delta P/P_1$=0.45 and denotes the
boundary between close and distant resolved peaks. Some objects are
tagged as both close and distant peaks (\eg, there are at least 3
periods in these objects, two of which are close and a third of which
is not close). Pulsators ($\delta$ Scutis) are largely outliers.  The
linear feature in the lower right predominantly becomes the slow
sequence found in the $P$ vs.~\vmk\ plot. The clump of sources near
$P\sim$0.4 and  $\Delta P/P_1\sim$0.3 becomes the M star fast sequence
in the $P$ vs.~\vmk\ plot. The lack of sources below the linear
feature is largely an observational bias.}
\label{fig:deltap}
\end{figure}

As described in Sec.~\ref{sec:resolvedmultifreq} above, for stars with
at least two periods, we calculated a metric, $\Delta P/P_1$, which is
the difference between the closest two periods in the periodogram,
divided by the primary period.  Figure~\ref{fig:deltap} shows the
relationship between this metric and period. The division between
close and distant peaks at $\Delta P/P_1$=0.45 can be seen there;
there are some stars with both close and distant peaks. Pulsators (the
$\delta$ Scutis) are often outliers in this diagram. (The possible
$\gamma$ Dors, having periods comparable to the rotation periods, are
not outliers in this diagram.) 

The linear feature in the lower right of Fig.~\ref{fig:deltap} is
composed of beaters and complex peaks, \eg, likely attributable to
differential rotation (or spot/spot group evolution).  Most shape
changers and double-dip stars are single-period stars, so few of them
can appear in this diagram, but those that do are found throughout
this diagram. While there is scatter, most of the stars in the linear
feature in the lower right of Fig.~\ref{fig:deltap} become the slow
sequence in the $P$ vs.~\vmk\ plot, and most of the multi-period stars
in the slow sequence in the $P$ vs.~\vmk\ plot populate the linear
feature in the lower right of Fig.~\ref{fig:deltap}. 

The concentration of sources near $P\sim$0.4 and  $\Delta
P/P_1\sim$0.3 in Fig.~\ref{fig:deltap} largely populates the M star
fast sequence in the $P$ vs.~\vmk\ plot. These fast rotating M
stars are, as we have seen above, likely to be solid body rotators.
Thus, these stars for which we can derive a second period are likely
to be binaries. They lie in a different portion of
Fig.~\ref{fig:deltap} than the stars likely to be differentially
rotating.  For completeness, we note that a few of the fast-rotating K
stars are also located in this $P\sim$0.4 and  $\Delta P/P_1\sim$0.3
clump.

The two most likely physical explanations for the \vmk$<$4 stars with
close, resolved peaks are (a) spots at two widely separated
longitudes, with significant evolution in spot size and/or shape  over
the K2 campaign, or (b) two or more spots/spot groups at different
latitudes for a star with significant latitudinal differential
rotation.  In the latter scenario, within this linear feature, the
faster the rotation rate is, the closer the peaks are, suggesting
weaker differential rotation in faster rotators -- consistent with the
fastest rotators (those that do not appear in this plot because they
do not have multiple peaks) rotating as solid bodies.

The surface of the Sun rotates differentially, with a relative shear
$\alpha$ (where $\alpha \Omega_{\rm eq} = (\Omega_{\rm eq} -
\Omega_{\rm pole}))$ of 0.2; that is, the pole rotates 20\% more
slowly than the equator.  If one assumes that the full range (pole to
equator) could be measurable in these LCs, $\Delta P/P_1 \sim 0.5$; if
one takes only the range over which sunspots are found, $\Delta
P/P_1 \sim 0.2$.  So, the Sun in Fig.~\ref{fig:deltap} lies more or
less along the extension of the slowly-rotating sequence of stars.
Reinhold, Reiners \& Basri (2013) have analysed data from the original
Kepler field for active low mass stars, and also find a good
correlation between  $\alpha$ and rotation period, though with much
more scatter than we see in the Pleiades (their sample of field stars
is much less homogeneous in age and metallicity than our sample).  

Reinhold \etal\ argue that the sloping lower bound to the points in
their figure is primarily an observational bias -- they simply could
not resolve two peaks in the LS periodogram if those two periods would
place them below that lower bound.   The lower bound to the linear
feature in Fig.~\ref{fig:deltap} is probably also largely an
observational bias for our data. We created a grid of synthetic models
with known, noise-free, sinusoidal periods of comparable amplitude to
test the limits of our approach to resolve the constituent periods.
Under these conditions, it becomes harder to resolve periods below the
linear feature. Recovery of the two periods is not precluded, but it
becomes less likely, as the relative amplitudes and phasing of the
constituent periods becomes more important. 

However, no such detection bias exists immediately above the linear
feature in our plot, from which we infer that for a coeval, young
population there is a well-defined degree of differential rotation at
a given mass, at least once the star is part of the slowly-rotating
sequence.  (We note, however, the role of spot/spot group evolution
may also be important for interpretation of these LCs.)

Recently, Balona \& Abedigamba (2016) find a similar relationship for
differentially rotating stars in the main Kepler field. They find that
$\Delta \Omega / \Omega$ as a function of $\Omega$ decreases
sharply with rotation rate; transforming Fig~\ref{fig:deltap} to that
parameter space finds that our linear feature is consistent with their
derived relationship.

\clearpage

\section{Conclusions}
\label{sec:concl}

We have continued our analysis of the K2 Pleiades lightcurves, finding
complicated multi-period behavior that we have grouped into
categories.  About 24\% of the sample has multiple, real frequencies
in the periodogram, sometimes manifesting as obvious beating in the
LCs. These LCs can fall into categories of those having complex and/or
structured periodogram peaks, unresolved multiple periods, and
resolved multiple periods. These are likely due to latitudinal
differential rotation and/or spot/spot group evolution. About 13\% of
the sample seems to have one period in the power spectrum, but the
light curve is seen to undergo substantial changes over the K2
campaign. These may be cases where the spot/spot group evolution is
slow over the 72 day K2 campaign, or cases in which the latitudinal
differential rotation is so weak that 72 days is not enough time to
separate the nearly identical periods.   About 1\% of the sample are
multi-period because they are likely $\delta$ Scuti pulsators. About
12\% of the sample have double-dipped or (double-humped) light curves
that redistribute power into peaks in the power spectrum other than
the main period. These light curves are probably a result of spots on
well-separated longitudes on the star.  In 6 cases, the light curves
have stable, shallow, angular patterns in the phased LC that affect a
relatively small fraction of the phase coverage. These shallow flux
dips may be due to transits or eclipses of orbiting clumps or clouds
near the Keplerian co-rotation radius. 

There are correlations of these broad categories with location in the
$P$ vs.~\vmk\ diagram. The slow sequence is dominated by complex
and/or structured peaks, unresolved multiple periods, and resolved
multiple periods; latitudinal differential rotation is likely to be happening in
these higher-mass stars. The fast sequence is dominated by
single-period stars; these are likely to be rotating as solid bodies.
The transition between the fast and slow sequence is dominated by the
shape changing LCs. Multiple period identifications among the lower
mass stars are likely to be binaries. Some multiple period
identifications among the higher mass stars may be multiples or even
pulsation, but the $P$ we have chosen is the one we believe to be
closest to the rotation period (that for the primary at least), so
this ambiguity is unlikely to affect our results.

For those stars where we can detect at least two periods, we can
calculate $\Delta P/P_1$, which is the difference between the closest
two periods in the periodogram, divided by the primary period. In the
plot of $\Delta P/P_1$ vs.~$P_1$, there is a striking linear feature 
in the lower right. It is composed of stars whose LCs are categorized
as beaters and complex peaks, and
they become the slow sequence in $P$ vs.~\vmk. While non-detection of
periods below this linear feature is likely an observational bias, no
such bias affects the distribution above the linear feature. Given
that, there is a well-defined correlation between the degreed of
latitudinal differential rotation and period for our sample.

We continue discussion of these results in Paper III (Stauffer \etal\
2016), which speculates about the origin and evolution of the period
distribution in the Pleiades.

\clearpage

\acknowledgments

We thank R.~Stern and T.~David for helpful comments on draft
manuscripts. ACC acknowledges support from STFC grant ST/M001296/1.

Some of the data presented in this paper were obtained from the
Mikulski Archive for Space Telescopes (MAST). Support for MAST for
non-HST data is provided by the NASA Office of Space Science via grant
NNX09AF08G and by other grants and contracts. This paper includes data
collected by the Kepler mission. Funding for the Kepler mission is
provided by the NASA Science Mission directorate. 

This research has made use of the NASA/IPAC Infrared Science Archive
(IRSA), which is operated by the Jet Propulsion Laboratory, California
Institute of Technology, under contract with the National Aeronautics
and Space Administration.    This research has made use of NASA's
Astrophysics Data System (ADS) Abstract Service, and of the SIMBAD
database, operated at CDS, Strasbourg, France.  This research has made
use of data products from the Two Micron All-Sky Survey (2MASS), which
is a joint project of the University of Massachusetts and the Infrared
Processing and Analysis Center, funded by the National Aeronautics and
Space Administration and the National Science Foundation. The 2MASS
data are served by the NASA/IPAC Infrared Science Archive, which is
operated by the Jet Propulsion Laboratory, California Institute of
Technology, under contract with the National Aeronautics and Space
Administration. This publication makes use of data products from the
Wide-field Infrared Survey Explorer, which is a joint project of the
University of California, Los Angeles, and the Jet Propulsion
Laboratory/California Institute of Technology, funded by the National
Aeronautics and Space Administration. 

\facility{Kepler} \facility{K2} \facility{Spitzer}
\facility{2MASS} 
facility{IRSA} facility{NASA Exoplanet Archive}
facility{Simbad} facility{Vizier}

\end{document}